\documentclass[10pt,journal,compsoc,onecolumn,english]{IEEEtran}
\ifCLASSINFOpdf
\else
\fi

\usepackage[T1]{fontenc}
\usepackage[utf8]{luainputenc}
\usepackage{array}
\usepackage{url}
\usepackage{multirow}
\usepackage{amsmath}
\usepackage{amssymb}
\usepackage{stmaryrd}
\usepackage{graphicx}
\usepackage{color}
\usepackage{xspace}
\usepackage[numbers]{natbib}

\definecolor{ScarletRed}{rgb}{0.80,0.00,0.00}

\providecommand{\tabularnewline}{\\}

\usepackage{braket}
\newcolumntype{L}[1]{>{\raggedright\let\newline\\\arraybackslash\hspace{0pt}}m{#1}} \newcolumntype{C}[1]{>{\centering\let\newline\\\arraybackslash\hspace{0pt}}m{#1}} \newcolumntype{R}[1]{>{\raggedleft\let\newline\\\arraybackslash\hspace{0pt}}m{#1}}
\usepackage{diagbox} 
\usepackage{tabularx}
\renewcommand{\arraystretch}{1}

\usepackage{array,graphicx}
\usepackage{booktabs}
\usepackage{rotating}
\newcommand*\rot{\rotatebox{90}}
\usepackage{amsmath,lipsum}

\usepackage{babel}
\usepackage{listings}

\begin{document}
%
\title{A Systematic Study of Cross-Project Defect Prediction With Meta-Learning}
%
%
%
%

\author{Faimison~Porto, Leandro~Minku, Emilia~Mendes, Adenilso~Simao
\IEEEcompsocitemizethanks{\IEEEcompsocthanksitem F. Porto and A. Simao are with the Institute of Math. and Computer Sciences (ICMC), University of Sao Paulo, Sao Carlos, Brazil.\protect\\
E-mail: \{faimison,adenilso\}@icmc.usp.br}
\IEEEcompsocitemizethanks{\IEEEcompsocthanksitem L. Minku is with the Department
of Informatics, University of Leicester, University Road, Leicester, LE1 7RH, UK.\protect\\
E-mail: leandro.minku@leicester.ac.uk}
\IEEEcompsocitemizethanks{\IEEEcompsocthanksitem E. Mendes is with the Department of Computer Science and Engineering, Faculty of Computing, Blekinge Institute of Technology, SE-37179, Karlskrona, Sweden; and M3S (Empirical Software Engineering in Software, Systems and Services) research unit, Faculty of Information Technology and Electrical Engineering, University of Oulu, P.O. Box 3000, 90014, Finland.\protect\\
E-mail: emilia.mendes@\{bth.se, oulu.fi\}}
\thanks{}}

%
%

\markboth{University of Birmingham, School of Computer Science Research Reports, CSR-18-01, ISSN 0962-3671}%
{Porto \MakeLowercase{\textit{et al.}}: A Systematic Study of Cross-Project Defect Prediction With Meta-Learning}
%



\IEEEtitleabstractindextext{%
\begin{abstract}
The prediction of defects in a target project based on data from external projects is called Cross-Project Defect Prediction (CPDP). Several methods have been proposed to improve the predictive performance of CPDP models. However, there is a lack of comparison among state-of-the-art methods. Moreover, previous work has shown that the most suitable method for a project can vary according to the project being predicted. This makes the choice of which method to use difficult. We provide an extensive experimental comparison of 31 CPDP methods derived from state-of-the-art approaches, applied to 47 versions of 15 open source software projects. Four methods stood out as presenting the best performances across datasets. However, the most suitable among these methods still varies according to the project being predicted. Therefore, we propose and evaluate a meta-learning solution designed to automatically select and recommend the most suitable CPDP method for a project. Our results show that the meta-learning solution is able to learn from previous experiences and recommend suitable methods dynamically. When compared to the base methods, however, the proposed solution presented minor difference of performance. These results provide valuable knowledge about the possibilities and limitations of a meta-learning solution applied for CPDP.
\end{abstract}

}

\maketitle

\IEEEdisplaynontitleabstractindextext

%
\IEEEpeerreviewmaketitle

\IEEEraisesectionheading{\section{Introduction}\label{sec:introduction}}

\IEEEPARstart{S}{oftware} testing activities are crucial quality assurance activities in a software development process. However, these activities are known to be expensive. In a scenario with limited resources (e.g., financial, time, manpower), a proper management of the available resources is necessary. In this context, software defect prediction models can be used to identify which parts of a software are more defect-prone and should thus receive more attention \cite{Malhotra2015}. 

The predictive power of a model is driven by the machine learning
technique employed and also to the quality of the training 
data. Usually, defect prediction models are investigated in the literature
using a within-project context that assumes the existence of previous
defect data for a given project \cite{Herbold2016}. This approach
is called Within-Project Defect Prediction (WPDP). However, in practice,
not all companies maintain clear historical defect data records or
sufficient data from previous versions of a project \cite{Kitchenham2007}. For
example, a software project in its first release has no historical
defect data. In such cases, an alternative approach is to assemble
a training set composed of known external projects. This approach,
named Cross-Project Defect Prediction (CPDP), has in recent years
attracted the interest of researchers due to its broader applicability
\cite{Gunarathna2016}. 

Although the CPDP principle is attractive, the predictive performance
has been shown to be a limiting factor, where poor predictive power
is commonly credited to the heterogeneity present on the data employed
\cite{Ma2012}. \citet{Zimmermann2009}, for example, show that
only 3.4\% of 622 pairs of projects ($\text{\emph{project}}_{1}$,
$\text{\emph{project}}_{2}$) presented adequate predictive power
when $\text{\emph{project}}_{1}$ was predicted based on a model trained
with $\text{\emph{project}}_{2}$. 

Other studies in the literature focused
on solutions to improve the predictive performance of CPDP models. 
Herein we emphasize transfer learning solutions \cite{Pan2010}.
Such approach is characterized by the use of some knowledge about
the target domain in order to approximate the different distributions
of source and target data. Different strategies are explored in this
context, such as the transformation of data \cite{Watanabe2008,Cruz2009},
the filtering of relevant examples or projects \cite{Turhan2009,Herbold2013,He2013},
or the different weighting of relevant training examples \cite{Ma2012}.
In addition, most of these methods can be associated to different machine
learning algorithms, composing variations with different predictive performances.
Knowledge about which methods are typically best could assist
practitioners select a method to use. However, to the best of our knowledge, no comprehensive comparison between such transfer learning solution methods has been published to date. Existing comparative analyses
are restricted to a few methods \cite{Ma2012,He2013,Herbold2016} or do not provide a uniform experimental
setup \cite{Gunarathna2016}.  

As part of the first contribution of this study, we provide an extensive
experimental comparison of 31 CPDP methods applied to 47 versions
of 15 open source software projects. The 31 methods derived from the
combination of six state-of-the-art transfer learning solutions with
the five most frequently used classifiers in the defect prediction
literature. We designed a uniform experimental setup to mitigate potential
sources of bias that could accidentally lead to conclusion instability
\cite{Menzies2012,Song2013,Shepperd2014,Tantithamthavorn2016}. In
this experimental comparison, we investigated which methods perform
better across datasets. Based upon the results obtained, we identified
four CPDP methods with best performance in general. We also investigated
whether these four methods performed better for the same datasets.
This answer can be helpful considering the different software domains
included in the investigated collection of datasets. Results showed
that even though these four methods performed best across datasets
the most suitable method for a
project varied according to the project (or dataset) being predicted.

Such result is also found in the broader machine learning research
area. The \emph{no free lunch} theorem states that no machine learning
algorithm is suitable for all domains \cite{Wolpert2002}. This statement
led us to also investigate: ``which is the most suitable solution
for a given domain?''. This task is named algorithm recommendation
\cite{Mantovani2015,dasDores2016,Parmezan2017}, which is part of
the general meta-learning literature \cite{Lemke2015,Sammut2017}.
A meta-learning model is characterized by its capacity of learning
from previous experiences and to adapt its bias dynamically conforming
to the target domain \cite{Sammut2017}. In the context of this study,
a meta-learning model should be able to recommend a suitable CPDP
method according to the characteristics of the software being predicted. 

Therefore, in this paper there is a second main contribution, which
is to propose and evaluate a meta-learning solution specifically designed
for the CPDP context. The particularities of this context are carefully
investigated and included in a general meta-learning architecture
\cite{Kalousis2002}. Such particularities include the unsupervised
characterization of datasets \cite{Santos2016}. Traditionally, such
characterization is given by supervised meta-attributes, which assumes
the existence of information about the target attribute \cite{Brazdil1994,Reif2014}.
However, in the CPDP context this information may not be available
(e.g., a project in its first release). Another particularity of the
proposed solution relates to the multi-label context \cite{Tsoumakas2007}.
The results obtained as part of our first contribution showed that
more than one CPDP method can achieve the best performance for a project,
which characterizes a multi-label learning task. These particularities
differentiate the proposed solution from previous studies, such as
the meta-learning solution proposed by \citet{dasDores2016} and \citet{Nucci2017},
designed for the WPDP context.

We consider two distinct meta-attributes sets for the characterization
of datasets. Both are evaluated in two different levels - meta- and
base-level. In the meta-level, we evaluate the learning capacity of
each meta-attributes set in relation to a baseline. In the base-level,
we evaluate the general performance of the proposed solution in relation
to the four best CPDP methods applied individually.

In general, we aim to answer the following research questions encompassing
both contributions:
\begin{itemize}
\item \textbf{RQ1: }\emph{Which CPDP methods perform better across datasets? }

We managed to identify the four best-ranked CPDP methods across datasets
in terms of AUC.

\item \textbf{RQ2: }\emph{Do the best CPDP methods perform better for the
same datasets? }

These four methods presented best performances for distinct groups
of datasets, evidencing that the most suitable method for a project
depends on the project being predicted. These results accredit the
investigation of new resources to assist the task of deciding which
methods are most suitable for a given project.

\item \textbf{RQ3: }\emph{To what extent can meta-learning help us to select
the most suitable CPDP method for a given dataset? }
\item \textbf{RQ3.1: }\emph{Does the meta-learner learn? (Meta-level)}

The meta-learner presented a better performance in relation to the
four evaluated CPDP methods considering the frequency of predicted
best solutions. In other words, the meta-learner was able to predict
the best solution for a larger amount of project versions. This result
indicates a proper predictive power of the proposed solution in the
meta-level.

\item \textbf{RQ3.2: }\emph{How does the meta-learner perform across datasets?
(Base-level)}

Even though the meta-learner presented the higher mean value of AUC,
it did not present significant difference in relation to the best-ranked
base method.

\end{itemize}
The remainder of this paper is organized as follows. In Section \ref{sec:Cross-project-defect-prediction}
we present the necessary background and related work. In Section \ref{sec:CPDP-Methods-Performance}
we present an extensive comparison of performance for the state-of-the-art
CPDP methods. In Section \ref{sec:Meta-learning-for-CPDP} we propose
a meta-learning architecture applied for CPDP and evaluate its performance.
In Section \ref{sec:Threats-to-Validity} we present the threats to
validity. In Section \ref{sec:Conclusions} we present the general
conclusions and comments on future work.

\section{Background and Related Work\label{sec:Cross-project-defect-prediction}}

\subsection{Software Defect Prediction\label{sub:Defect-Prediction}}

The success of a software defect prediction model depends on two main
factors: 1) building an adequate training set; and 2) applying a suitable
machine learning technique. 

In a classification task, the training set can be represented by a
table of elements (see Table \ref{tab:Classification-training-dataset.}).
In this table, each row represents an example (software part or module);
the independent variables (or attributes) represent the characteristics
of each example; and the dependent variable (class or target attribute)
represents the binary class - defective or not defective.

\begin{table}[!t]
\centering
\caption{\label{tab:Classification-training-dataset.}Classification Training
Set.}
\footnotesize
\begin{tabular}{cccccc}
\hline 
\multirow{2}{*}{Examples} & \multicolumn{4}{c}{Attributes} & \multirow{2}{*}{Class}\tabularnewline
\cline{2-5} 
 & $A_{1}$ & $A_{2}$ & $\dots$ & $A_{m}$ & \tabularnewline
\hline 
$E_{1}$ & $x{}_{11}$ & $x_{12}$ & $\dots$ & $x_{1m}$ & $c_{1}$\tabularnewline
$E_{2}$ & $x_{21}$ & $x_{22}$ & $\dots$ & $x_{2m}$ & $c_{2}$\tabularnewline
$\vdots$ & $\vdots$ & $\vdots$ & $\ddots$ & $\vdots$ & $\vdots$\tabularnewline
$E_{n}$ & $x_{n1}$ & $x_{n2}$ & $\dots$ & $x_{nm}$ & $c_{n}$\tabularnewline
\hline 
\end{tabular}
\end{table}

Much effort has been spent in order to find effective independent
variables, i.e., software properties able to provide relevant information
for the learning task. Seminal studies were presented in \citet{Basili1996},
where the authors investigated the use of object-oriented metrics.
\citet{Ostrand2004} proposed the use of code metrics and historical
defect information for defect prediction in industrial large-scale
systems. \citet{Zimmermann2008} proposed the use of social network
analysis metrics, extracted from the software dependency graph. Further,
other metrics were also investigated, such as developer-related metrics
\cite{Pinzger2008}, organizational metrics \cite{Nagappan2008},
process metrics \cite{Hassan2009}, change related metrics \cite{Herzig2013},
antipatterns related metrics \cite{Taba2013}, to name a few. Herein
we use the code metrics set investigated in \citet{Jureczko11} because
such metrics set has been reported as effective for defect prediction
models \cite{Jureczko11,Herbold2013,Madeyski2014,Ghotra2015}, and
can be automatically extracted from a project’s source code. This
latter characteristic is important due to two factors: 1) no historical
or additional data is required (e.g., process metrics or change-related
metrics); and 2) any software project with available source code can
be used as input for the prediction model. 

There are also additional studies focusing upon the performance of
classification algorithms applied to defect prediction. \citet{Lessmann2008}
compared the performances of 22 classifiers over 10 public domain
datasets from the NASA Metrics Data Repository (MDP). Their results
show no statistically significant differences between the top 17 classifiers.
Later, \citet{Ghotra2015} revisited \citeauthor{Lessmann2008}’s
work considering two different collections of data: the cleaned NASA
MDP; and the open source PROMISE\footnote{\url{http://openscience.us/repo/}}
corpus. Opposite to the previous results, \citeauthor{Ghotra2015}
concluded that the used classification technique had a significant
impact on the performance of defect prediction models. These contrasting
results reinforce the conclusion instability (discussed in Section
\ref{sub:Conclusion-Instability}), as well as highlight the influence
of noisy data upon results. In this study, we employed data from the
latest version of the open source PROMISE data corpus, provided by
\citet{Madeyski2014} (presented in Section \ref{sub:Software-Projects}).

Finally, \citet{Malhotra2015} presents a systematic literature review
(SLR) on machine learning techniques for software defect prediction,
based on 64 primary studies (from 1991 to 2013). This SLR identified
the five most frequently used classifiers: Naive Bayes; Random Forest;
Support Vector Machines; Multilayer Perceptron; and C4.5. A better
description of each classifier is provided in Appendix \ref{tab:five-classifiers}.
These five classifiers are evaluated herein, combined with transfer
learning solutions, since their different algorithms can lead to different
performance.

All the studies abovementioned employed the same methodology, called
within-project defect prediction (WPDP) \cite{Herbold2016}. In this
methodology, the training set includes examples from the same software.
Therefore, the prediction of a software defect is based on this software’s
previous versions. However, an appropriate amount of historical defect
data may not be available. An alternative is to compose the training
set with external known projects, as discussed next.

\subsection{Cross-project Defect Prediction\label{sub:Cross-project-Defect-Prediction}}

Within the context of software defect prediction, the dependent variable,
i.e., what we aim to predict, also has an important role. In fact,
learning is only possible when defect data is available. However,
in practice, not all software companies maintain clear records of
historical defect data or sufficient data from previous projects \cite{Kitchenham2007}.
In this case, the training set can be composed by external projects
with known defect information. This approach is called Cross-project
Defect Prediction (CPDP) \cite{Gunarathna2016}. The positive aspect
of such approach is that it tackles the lack of historical defect
data; however, on the negative side, it introduces heterogeneity on
data, which may decrease the efficiency of defect prediction models
\cite{Zimmermann2009}. 

In a recent systematic literature review, \citet{Gunarathna2016}
highlighted the contributions of 46 primary studies on CPDP from 2002
to 2015. Part of these works focus on the feasibility of CPDP. \citet{Briand2002}
conducted one of the earliest studies on this topic and verified that
CPDP models outperform random models in the studied case. \citet{Zimmermann2009}
conducted a large scale experiment in which they evaluated 622 pairs
of cross-project models ($\text{\emph{project}}_{1}$, $\text{\emph{project}}_{2}$).
They found that a model trained with $\text{\emph{project}}_{1}$
was able to predict $\text{\emph{project}}_{2}$ with good performance
in only $3.4\%$ of pairs. 

Other studies are focused on possible solutions to mitigate the heterogeneity
of the data and to improve the learning capacity of CPDP models. Among
all solutions, we highlight the transfer learning solutions\footnote{The context of this work fits in a special case of transfer learning,
called domain adaptation \cite{Pan2010}. In this case, source (external
projects) and target (project to be predicted) domains may be different
while sharing the same learning task. In addition, the target domain
may be unlabelled (e.g., the first release of a software project,
with no historical defect data).} \cite{Pan2010}. In this approach, the solutions use some knowledge
about the target domain in order to approximate the different distributions
of source and target data.

At least three different strategies can be identified among the transfer
learning solutions: transformation of data \cite{Watanabe2008,Cruz2009,Nam2013,Zhang2014};
filtering a subset of the training data \cite{Turhan2009,Jureczko2010,Herbold2013,He2013};
and weighting the training data according to the target data \cite{Ma2012}.

\citet{Watanabe2008} proposed to standardize the target data based
on the source mean values. \citet{Cruz2009} put forward a similar
approach to the one by \citeauthor{Watanabe2008}, however based on
the median values, associated with power transformation. \citet{Nam2013}
suggested to project both source and target data in a common attribute
space by means of TCA (Transfer Component Analysis). Moreover, they
investigated the impact of min/max and Z-score standardizations applied
over source and target data. \citet{Zhang2014} offered a universal
defect prediction model based on clusters created using the project
context.

\citet{Turhan2009} suggested the use of the k-nearest neighbour (KNN)
filter to select the most similar training examples in relation to
the target examples. \citet{Jureczko2010} recommended to perform
clustering in order to identify the most suitable training projects
for a target project. \citet{Herbold2013} put forward the use of
the k-nearest neighbour filter but to select the most similar projects
(instead of individual examples). \citet{He2013} suggested the selection
of the most similar projects but considering the separability of each
project in comparison to the target data. Moreover, they also suggested
the use of the separability strategy to filter unstable attributes.
Lastly, they applied an ensemble approach over the selected projects
with filtered attributes.

\citet{Ma2012} used the idea of data gravitation to prioritize and
weight the training data for the Naive Bayes classifier based on its
similarity to the target data.

To the best of our knowledge, no extensive comparison of transfer
learning methods were conducted in the CPDP context. Commonly, authors
compared their new solutions either with the KNN filter (e.g., \cite{Ma2012,He2013,Herbold2016})
or with some baseline (e.g., \cite{Herbold2013,Nam2013,Herbold2016}).
The baseline, usually, is defined as the direct application of a machine
learning technique, with no CPDP treatment.

In the systematic literature review presented in \citet{Gunarathna2016},
the authors compared the performance of CPDP methods based on the
information provided by each published work. However, the considered
studies diverge in terms of data, classifiers, and performance measures.
The absence of a uniform experimental setup precludes concrete comparisons
of performances between studies on a large scale. This issue is better
discussed in the next section.

\subsection{Conclusion Instability\label{sub:Conclusion-Instability}}

Defect prediction models assist testers on prioritizing test resources.
Yet, there is another important decision in this process: which model
is the most suitable for a specific domain? The answer to this question
is not straightforward for two main reasons. First, as stated in \citet{Shepperd2014},
no single defect prediction technique dominates. This statement has
support in the theorem known as \emph{no free lunch}, which says that
there is no machine learning algorithm  suitable for all domains
\cite{Wolpert2002}. Second, because of the \emph{conclusion instability}
found in the experimental software engineering \cite{Menzies2012}.
Not rarely, different studies on the same subject produce conflicting
conclusions. For example, \citet{Kitchenham2007} reviewed empirical
studies applied to effort estimation in which local data are evaluated
in relation to data imported from other organizations. From the seven
reviewed studies, three concluded that imported data are not worse
than local data, while four studies concluded that imported data are
significantly worse than local data. 

Part of this conclusion instability is credited to the bias produced
by the different research groups \cite{Shepperd2014}. However, in
a recent study, \citet{Tantithamthavorn2016} argue that this bias
is actually due to the strong tendency of a research group to reuse
experimental resources such as datasets and metrics families. 

\citet{Menzies2012} also credit the conclusion instability to the
variance between similar experiments. This variance is related to
the internal resources and procedures used in each experimentation.
One of the suggestions proposed by \citet{Menzies2012} to reduce
the conclusion instability is the use of a uniform experimental setup.
In the context of CPDP, several factors can influence the performance
analysis, such as:
\begin{itemize}
\item The collection of software projects used as training and test sets:
different data repositories are used in the literature of CPDP, such
as PROMISE, Apache, Eclipse, and NASA MDP \cite{Turhan2009,Herbold2013,He2015};
\item The independent variables set: some of the data repositories mentioned
above provide datasets with different metrics sets. The metrics set
is an important source of performance variability \cite{Tantithamthavorn2016};
\item The modelling technique: although some solutions are intrinsically
associated to some classifier \cite{Ma2012}, other transfer learning
solutions are independent of the learning algorithm \cite{Watanabe2008,Turhan2009,Herbold2013}
and their performance can be influenced by the applied classifier;
\item The cross-project configuration: one-to-one, when only one dataset
(or software project) is used for training \cite{Zimmermann2009};
or many-to-one, when multiple projects compose the training set \cite{Turhan2009};
\item Data preprocessing: different methods can be applied in the preprocessing
step, such as normalization, standardization, attribute selection,
or data transformation \cite{Maimon2010}. These methods modify the
value range, distribution and scale of data, interfering directly
on the CPDP model performance \cite{Keung2013,Nam2013}; and
\item Performance measure: some traditional metrics (e.g., precision, recall,
f-measure) are sensitive to the threshold that separates a defective
from a non-defective example. Other metrics (e.g., AUC), however,
are unrestricted to this threshold. Furthermore, different metrics
focus on different aspects of the performance. Thus, a proper performance
comparison demands the use of equivalent measures.
\end{itemize}
Considering each of the bias factors abovementioned, in Section \ref{sec:CPDP-Methods-Performance}
we define a uniform experimental setup for the performance comparison
of 31 CPDP methods\footnote{In this work, the term CPDP method refers to the entire process (transfer
learning solution + classifier) which leads to the prediction model.} derived from the combination of six state-of-the-art transfer learning
solutions with the five most popular classifiers for defect prediction.

In order to mitigate conclusion instability, \citet{Menzies2012}
also suggest the investigation of approaches able to learn the properties
of a particular domain and, as a result, suggest a suitable learner.
This suggestion is strongly related to the meta-learning architecture
proposed and evaluated in Section \ref{sec:Meta-learning-for-CPDP}.
We discuss the background of this proposal in the next section.

\subsection{Meta-learning for Algorithm Recommendation\label{sub:Meta-learning}}

Considering the \emph{no free lunch} theorem and the concept of \emph{conclusion
instability }discussed above, automated resources to assist practitioners
in the difficult task of choosing a suitable model for a specific
domain are desirable. This task is related to the algorithm recommendation
task investigated in the meta-learning literature \cite{Lemke2015}.
The learning capacity of a traditional model (here identified as base-learner)
is limited to its inductive bias \cite{Mitchell1997}. A meta-learning
model, however, is characterized by its capacity of learning from
previous experiences and adapting its inductive bias dynamically according
to the target domain \cite{Sammut2017}. In other words, a meta-learning
model is designed to learn in which conditions a given solution is
more suitable to be applied to, when compared to all conditions.

The main challenge in the algorithm recommendation task is to discover
the relationship between measurable features of the problem and the
performance of different solutions \cite{Rice1976}. In the literature,
this task is approached as a typical classification problem, although
at a meta-level \cite{Kalousis2002,Ali2006,Sammut2017}.

Similar to the traditional learning task, the success of a meta-model
depends on two main factors: 1) building an accurate meta-data; and
2) applying a suitable machine learning algorithm to compose the meta-learner.
The meta-data is represented by a table of meta-examples. Each meta-example
represents a dataset (i.e., a previous learning task experience);
the independent variables (or meta-attributes) characterize each dataset;
and the dependent variable (or meta-target) represents the goal of
the meta-learning task. 

Different goals have been explored in the literature for both classification
and regression tasks. In classification tasks, the meta-target is
commonly associated to a label representing the solution with highest
performance (or lower cost). Examples of meta-learning classification
tasks are: to predict a suitable base-learner \cite{dasDores2016,Nucci2017};
to predict a suitable attribute selection method \cite{Parmezan2017};
to predict if the tuning of hyper-parameters for SVMs is beneficial
or not \cite{Mantovani2015}; among others \cite{Lemke2015}. In
regression tasks, each meta-target represents the continuous value
of performance (or cost) obtained by a specific solution. In this
context, the meta-learning task is commonly designed to produce a
rank of possible solutions \cite{Sammut2017}. The recommended order
can be useful either to guide the decision task \cite{Kanda2016}
or to compose ensemble solutions \cite{Cruz2017}. 

Another important aspect that is highly related to the goal of the
meta-learning task is the set of meta-attributes. Traditionally, they
are classified into five main categories \cite{Brazdil1994,Reif2014}:
general attributes, obtained directly from the properties of the dataset
(e.g., number of examples, number of attributes); statistical attributes,
obtained from statistical measures (e.g., correlation between attributes);
information-theoretic attributes, typically obtained from entropy
measures (e.g., class entropy, attribute entropy); model-based attributes,
extracted from internal properties of an applied model (e.g., the
width and height of a decision tree model); and landmarking attributes,
obtained from the resulting performance of simple classifiers (e.g.,
the accuracy obtained from a 10-fold cross-validation with 1-nearest
neighbour classifier). 

In the context of this study, for example, all datasets present the
same set of continuous attributes (see Section \ref{sub:Software-Projects}).
In addition, the target data may be unlabelled, as discussed in Section
\ref{sub:Cross-project-Defect-Prediction}. These two characteristics
make most of the traditional meta-attributes unsuitable to use, since
many of them are related to discrete attributes, to the dimensionality
of the dataset, or are label dependent - including information-theoretic,
model-based and landmarking attributes. 

A viable alternative is to characterize the datasets based on unsupervised
meta-attributes (non-dependent on the class attribute). However, this
alternative is little explored in the algorithm recommendation literature.
Although some of the traditional meta-attributes are not dependent
on the class attribute, they are commonly addressed associated with
label dependent meta-attributes. In a recent study, \citet{Santos2016}
proposed a set composed of 53 meta-attributes applied in the context
of active learning, including only unsupervised attributes such as
general attributes, statistical attributes and clustering based attributes.
Another related approach is to represent a dataset based on its distributional
characteristics (e.g., mean, maximum and standard deviation of all
attributes). Although this latter approach is not commonly applied
for algorithm recommendation tasks, it is already applied in the context
of CPDP with different purposes \cite{He2012,He2013,Herbold2013}.
For example, \citet{Herbold2013} uses the distributional characteristics
of datasets to filter the most similar projects to compose the final
training set. These two unsupervised approaches are explored and evaluated
in this study. A better description is provided in Section \ref{sub:Meta-features}.

In principle, any machine learning algorithm can be used at the meta-level.
However, a common aspect of meta-learning tasks is the scarce set
of training data \cite{Sammut2017}. This issue is mitigated in the
literature with lazy learning methods, such as k-nearest neighbour
classifier \cite{Lemke2015,Sammut2017}, although other types of
models have been successfully applied (e.g., neural networks, random
forest, ensembles) \cite{dasDores2016,Parmezan2017}. 

In \citet{dasDores2016}, the authors proposed a meta-learning framework
for algorithm recommendation in the context of WPDP. Their meta-learning
solution includes a meta-data composed of seven classifiers applied
over 71 distinct datasets. The datasets are characterized with traditional
meta-attributes, as described above. They evaluated two meta-learners:
a Random Forest model; and a majority voting ensemble, including all
seven classifiers. Their experiments reveal a better performance obtained
with the meta-learning solution across datasets in relation to each
of the seven classifiers applied individually. However, no statistical
analysis was reported. In \citet{Nucci2017}, the authors proposed
a different approach for the dynamic selection of classifiers in the
context of WPDP. Their solution differs from \citet{dasDores2016}
for two main reasons: 1) their solution considers a different set
of classifiers applied over 30 distinct datasets; and 2) their meta-data
is constructed in a different level of granularity, i.e., the meta-examples
do not represent software project versions (or datasets) but single
Java classes. In this way, the characterization of meta-examples are
based on the same independent variables used in the base-level. The
meta-learner considered in their work is also based on the Random
Forest algorithm. Their results are positive since the proposed solution
outperformed the evaluated stand-alone classifiers and a voting ensemble
technique.

In Section \ref{sec:Meta-learning-for-CPDP}, we propose and evaluate
a different meta-learning solution designed specifically for the CPDP
context. The particularities of this context lead to new issues, carefully
investigated in this study. Some of such issues relate to the unsupervised
meta-attributes set and the multi-label context (see Section \ref{sub:Multi-label-Classification}).
Furthermore, a series of other factors distinguish the proposed solution
from previous meta-models, including the collection of datasets, the
validation process, the performance measure, the attribute selection
procedure, and the statistical analysis.

\subsection{Multi-label Learning \label{sub:Multi-label-Classification}}

The meta-learning solution proposed in this study also configures
a multi-label learning task since each meta-example can be associated
to more than one suitable CPDP method. In this section we present
the main concepts related to this particular domain.

In the traditional \emph{single-label }classification task, as described
in Section \ref{sub:Defect-Prediction}, each example of a dataset
is associated with a single label $\lambda$ from a set of disjoint
labels $\mathcal{L}=\{\lambda_{1},...,\lambda_{c}\}$, where $c=|\mathcal{L}|$
and $c>1$. The learning task is called binary classification when
$c=2$ and multi-class classification when $c>2$. In a \emph{multi-label}
classification task, each example is associated with a set of labels
$L\subseteq\mathcal{L}$ \cite{Tsoumakas2007}. 

\begin{table}[!t]
\centering
\caption{\label{tab:Multi-label-classification-datas}Multi-Label Training
Set.}
\footnotesize
\begin{tabular}{ccccccccc}
\hline 
\multirow{2}{*}{Examples} & \multicolumn{4}{c}{Attributes} & \multicolumn{4}{c}{Labels (Classes)}\tabularnewline
\cline{2-9} 
 & $A_{1}$ & $A_{2}$ & $\dots$ & $A_{m}$ & $\lambda_{1}$ & $\lambda_{2}$ & $\dots$ & $\lambda_{c}$\tabularnewline
\hline 
$E_{1}$ & $x_{11}$ & $x_{12}$ & $\dots$ & $x_{1m}$ & $y_{11}$ & $y_{12}$ & $\dots$ & $y_{1c}$\tabularnewline
$E_{2}$ & $x_{21}$ & $x_{22}$ & $\dots$ & $x_{2m}$ & $y_{21}$ & $y_{22}$ & $\dots$ & $y_{2c}$\tabularnewline
$\vdots$ & $\vdots$ & $\vdots$ & $\ddots$ & $\vdots$ & $\vdots$ & $\vdots$ & $\ddots$ & $\vdots$\tabularnewline
$E_{n}$ & $x_{n1}$ & $x_{n2}$ & $\dots$ & $x_{nm}$ & $y_{n1}$ & $y_{n2}$ & $\dots$ & $y_{nc}$\tabularnewline
\hline 
\end{tabular}
\end{table}

Let $\mathbf{x_{i}}\in\mathbb{R}^{m\times1}$ be a real value example
vector, $\mathbf{y}_{\mathbf{i}}\in\{0,1\}^{c\times1}$ a label vector
for $\mathbf{x}_{\mathbf{i}}$, $n$ the number of training examples
and $\mathbf{y}_{\mathbf{ij}}$ the $j$th label of the $i$th example,
where $1\leq i\leq n$ and $1\leq j\leq c$. The element $\mathbf{y_{ij}}$
is $1$ (or $0$) when the $j$th label is relevant (or irrelevant)
for the $i$th example. We denote $L_{i}$ as the set of relevant
labels for the example $\mathbf{x}_{\mathbf{i}}$. A training set
is composed by the example matrix $X\in\mathbb{R}^{n\times m}$ and
the label matrix $Y\in\{0,1\}^{n\times c}$ (see Table \ref{tab:Multi-label-classification-datas}). 

Given a multi-label training set ($X$, $Y$), the learning task can
have two goals: classification or ranking \cite{Schapire2000}. A
multi-label classification model is defined by $H:\mathbb{R}^{m}\rightarrow\{0,1\}^{c}$.
A multi-label ranking model is defined by $F:\mathbb{R}^{m}\rightarrow\mathbb{R}^{c}$,
where the predicted value can be regarded as the confidence of relevance.
The ranking is given by ordering the resulting confidence of relevance.
We can also define a multi-label classifier $F$ that assigns a single
label for $\mathbf{x}_{\mathbf{i}}$ by setting $\text{arg max }F(\mathbf{x_{i}})$.
In other words, the predicted single label for $\mathbf{x}_{\mathbf{i}}$
is given by the top ranked confidence of relevance generated by $F$. 

Several methods have been proposed in the literature of multi-label
learning \cite{Maimon2010}. The existing methods can be grouped
into two main categories: 1) problem transformation, in which the
multi-label learning task is transformed into one or more single-label
learning tasks; and 2) algorithm adaptation, in which specific learning
algorithms are adapted to handle multi-label problems directly.

In this study we highlight a problem transformation method called
Binary Relevance (BR) \cite{Luaces2012}. In this method, the multi-label
problem is transformed into $c$ binary classifiers, one for each
different label in $\mathcal{L}$. Then, a dataset $D_{\lambda_{j}}$,
$1\leq j\leq c$, is created for each label $\lambda_{j}\in\mathcal{L}$
containing all examples of the original dataset. Each example $\mathbf{x_{i}}$
in $D_{\lambda_{j}}$ is labelled positively if $\lambda_{j}\in L_{i}$
and negatively otherwise. 

In this study, this method is appropriate due to three main reasons:
1) BR presents a competitive performance with respect to more complex
methods \cite{Luaces2012}; 2) this is a popular, simple and intuitive
method \cite{Maimon2010}; and 3) after data transformation, any
binary learning method can be taken as base learner \cite{Luaces2012}.
This characteristic enables us to construct the multi-label model
discussed in Section \ref{sub:Meta-model}.

\section{Performance Evaluation of CPDP Methods \label{sec:CPDP-Methods-Performance}}

In this section we evaluate the performance of $31$ CPDP methods
including six state-of-the-art transfer learning solutions and their
variations combined with the five most popular classification learners
in the literature of defect prediction. We provide a uniform experimental
setup based on $47$ versions of $15$ open source software projects. 

The contribution of this experimental analysis is twofold: 1) it provides
a comprehensive evaluation of performance for the state-of-the-art
CPDP methods, covering different classification algorithms and transfer
learning methods; and 2) the results show that, despite the fact that
a group of methods generally present better predictive performance
than other methods across datasets, the most suitable method can vary
according to the project being predicted - accrediting the meta-learning
solution proposed in Section \ref{sec:Meta-learning-for-CPDP}.

Below, we discuss the experimental setup and issues involved in this
experimentation. Next, we present and discuss the obtained results.
All the experiments and data analysis were implemented and conducted
with R\footnote{\url{https://www.r-project.org/}}.

\subsection{Experimental Setup}

\subsubsection{Software Projects\label{sub:Software-Projects}}

All the conducted experiments are based on 47 versions of 15 Java
open source projects, provided by \citet{Madeyski2014}. The authors
provide a link\footnote{\url{http://purl.org/MarianJureczko/MetricsRepo/}}
with detailed information about the software projects and the construction
of each dataset. Each example in a dataset represents a Java Object-Oriented
class (OO class). The dependent variable corresponds to the number
of defects found for each OO class. We converted the dependent variable
to a binary classification problem ($1$, for number of defects $>$
$0$; or $0$, otherwise) \cite{He2013}. Table \ref{tab:ap-projs}
list the number of examples, number of defects, and defect rate for
each of the analysed datasets.

\begin{table}[!t]
\centering
\caption{\label{tab:ap-projs}Summary of the Projects Characteristics. }
\begin{tabular}{lccc} \hline \textbf{Project}  & \textbf{\# Examples} & \textbf{\# Defective Examples} & \textbf{\% Defective Examples} \\ \hline ant-1.3           & 180                   & 20                              & 0.11                            \\ \hline ant-1.4           & 253                   & 39                              & 0.15                            \\ \hline ant-1.5           & 381                   & 32                              & 0.08                            \\ \hline ant-1.6           & 496                   & 92                              & 0.19                            \\ \hline ant-1.7           & 999                   & 166                             & 0.17                            \\ \hline camel-1.0         & 409                   & 13                              & 0.03                            \\ \hline camel-1.2         & 687                   & 209                             & 0.3                             \\ \hline camel-1.4         & 1009                  & 144                             & 0.14                            \\ \hline camel-1.6         & 1117                  & 184                             & 0.16                            \\ \hline ckjm-1.8          & 10                    & 5                               & 0.5                             \\ \hline forrest-0.7       & 31                    & 5                               & 0.16                            \\ \hline forrest-0.8       & 34                    & 2                               & 0.06                            \\ \hline ivy-1.1           & 131                   & 63                              & 0.48                            \\ \hline ivy-1.4           & 311                   & 16                              & 0.05                            \\ \hline ivy-2.0           & 459                   & 40                              & 0.09                            \\ \hline jedit-3.2.1       & 492                   & 90                              & 0.18                            \\ \hline jedit-4.0         & 579                   & 75                              & 0.13                            \\ \hline jedit-4.1         & 616                   & 79                              & 0.13                            \\ \hline jedit-4.2         & 764                   & 48                              & 0.06                            \\ \hline jedit-4.3         & 1049                  & 8                               & 0.01                            \\ \hline log4j-1.0         & 152                   & 34                              & 0.22                            \\ \hline log4j-1.1         & 122                   & 37                              & 0.3                             \\ \hline log4j-1.2         & 258                   & 187                             & 0.72                            \\ \hline lucene-2.0        & 276                   & 91                              & 0.33                            \\ \hline lucene-2.2        & 355                   & 141                             & 0.4                             \\ \hline lucene-2.4        & 497                   & 199                             & 0.4                             \\ \hline pbeans-1.0        & 32                    & 20                              & 0.62                            \\ \hline pbeans-2.0        & 63                    & 10                              & 0.16                            \\ \hline poi-1.5           & 237                   & 130                             & 0.55                            \\ \hline poi-2.0RC1        & 304                   & 36                              & 0.12                            \\ \hline poi-2.5.1         & 383                   & 223                             & 0.58                            \\ \hline poi-3.0           & 441                   & 257                             & 0.58                            \\ \hline synapse-1.0       & 158                   & 16                              & 0.1                             \\ \hline synapse-1.1       & 223                   & 60                              & 0.27                            \\ \hline synapse-1.2       & 257                   & 86                              & 0.33                            \\ \hline tomcat-6.0.389418 & 1052                  & 77                              & 0.07                            \\ \hline velocity-1.4      & 205                   & 132                             & 0.64                            \\ \hline velocity-1.5      & 225                   & 133                             & 0.59                            \\ \hline velocity-1.6.1    & 237                   & 76                              & 0.32                            \\ \hline xalan-2.4.0       & 818                   & 110                             & 0.13                            \\ \hline xalan-2.5.0       & 873                   & 365                             & 0.42                            \\ \hline xalan-2.6.0       & 869                   & 326                             & 0.38                            \\ \hline xalan-2.7.0       & 888                   & 732                             & 0.82                            \\ \hline xerces-1.2.0      & 426                   & 64                              & 0.15                            \\ \hline xerces-1.3.0      & 451                   & 69                              & 0.15                            \\ \hline xerces-1.4.4      & 573                   & 378                             & 0.66                            \\ \hline xerces-init       & 193                   & 67                              & 0.35                            \\ \hline \textbf{Total}    & \textbf{20575}        & \textbf{5386}                   & \textbf{0.26}                   \\ \hline \end{tabular}
\end{table}

The set of independent variables is composed of 20 code metrics and
include complexity metrics, C\&K (Chidamber and Kemerer) metrics,
and structural code metrics. This metrics set is briefly described
in Appendix \ref{tab:code_metrics}. These metrics have been
reported in the literature as good quality indicators, as discussed
in \citet{Jureczko2010}. In addition, they are numeric and can be
automatically extracted directly from the source code with the Ckjm\footnote{\url{http://gromit.iiar.pwr.wroc.pl/p_inf/ckjm}}
tool. Further details can be found in \citet{Jureczko2010}.

\subsubsection{Data Preprocessing\label{sub:Data-Preprocessing}}

Data transformation is a resource used to re-express data in a new
scale aiming to spread skewed curves equally among the batches \cite{Cruz2009}.
The result is a distribution near to the normal distribution assumption
with less variability, better symmetry, reduced number of outliers
and better effectiveness for data mining \cite{Cruz2009,Turhan2009}.
We applied a log transformation. Each numeric value is replaced by
its logarithm. Since some metrics present zero values which would
result on infinitive values, we applied a simple solution as follows:
x = log(x + 1) \cite{Cruz2009}. 

Log transformation provides a uniform data preprocessing for all projects
since it is not dependent on the current test and training sets. In
\citet{Nam2013} and \citet{Keung2013}, the authors discuss the considerable
impact on performance caused by different preprocessing resources.

\subsubsection{CPDP Methods\label{sub:CPDP-Methods}}

In order to carry out this experiment we selected six state-of-the-art
transfer learning solutions based on the benchmark proposed in \citet{Herbold2013}.
We re-implemented each solution in R following their respective original
publication instructions. We refer to each solution with the abbreviation
{[}year{]}{[}first author{]}, as described bellow:
\begin{itemize}
\item \emph{2008Watanabe~\cite{Watanabe2008}:} the authors proposed a
simple compensation method based on the mean value. Let $x_{ij}$
be the $j$th attribute of the test example $x_{i}$, and $mean(Tr_{j})$
and $mean(Te_{j})$ be the mean values of the $j$th attribute of
the training and test sets, respectively. The compensated test metric
value $x_{ij}'$ is given by:
\[
x_{ij}'=x_{ij}*[mean(Tr_{j})/mean(Te_{j})].
\]
The base-learner is then applied over the transformed data.
\item \emph{2009Cruz~\cite{Cruz2009}:} the authors proposed a similar
compensation method based on the median value, associated with a log
transformation (already defined in Section \ref{sub:Data-Preprocessing}).
Considering the notations defined above, the compensated test metric
value $x_{ij}'$ is given by:
\[
x_{ij}'=x_{ij}+[median(Tr_{j})-median(Te_{j})].
\]

\item \emph{2009Turhan~\cite{Turhan2009}:} also known as KNN filter or
Burak filter, this solution applies the k-nearest neighbour algorithm
to select the most similar examples based on the Euclidean distance.
For each test example, the solution searches for the $k$ most similar
training examples (from all available projects). The selected examples
compose the filtered training set. Duplicated examples are discarded.
As in the original proposal, we set $k=10$. 
\item \emph{2013He~\cite{He2013}:} this solution is composed of three
steps. First, the $N$ projects datasets most similar to the test
set are selected to compose the training data. Let $D_{train}^{i}$
be the dataset of project $i$ ($1<i<N$) and $D_{test}$ the test
set. In this step, a new dataset called $SAM_{i}$ with $2K$ examples
is constructed containing $K$ examples from $D_{train}^{i}$ and
$K$ examples from $D_{test}$. The dependent variable of $SAM_{i}$
is a binary label differentiating training examples from test examples.
The similarity between $D_{train}^{i}$ and $D_{test}$ is calculated
based on the accuracy (rate of correct predictions) in which a classifier
is able to differentiate (or separate) between training and test examples
in $SAM_{i}$. In this step we applied a Logistic Regression as the
separability classifier and set $N=10$, as suggested by the authors.
\\
Second, for each of the project datasets selected in the previous
step, the ratio of $FSS*100$ percent of unstable attributes are removed,
where $FSS$ is a predefined parameter. The same dataset $SAM_{i}$
is used to measure the information gain of each attribute in relation
to the dependent variable. The attributes with higher information
gains are considered more unstable and are removed. It is important
to note that each of the selected projects (step 1) generates a new
training set with, possibly, a distinct attributes subset (step 2).
We set $FSS=0.8$, as suggested by the authors. \\
Finally, the authors apply an ensemble strategy with majority voting
to combine the predictions given by the selected training sets (obtained
from steps 1 and 2). For each selected training set, a model is constructed
using a classifier. Then, for each test example, a prediction score
is generated for each constructed model. In the original proposal,
a threshold of $0.5$ is applied for each prediction score and the
majority voting defines the final binary prediction. \\
In our implementation, since we are working with the AUC measure (obtained
from the prediction score), we considered the average prediction score
obtained from the ensemble strategy instead of the majority voting.
\item \emph{2013Herbold~\cite{Herbold2013}: }in the original work, the
authors proposed and evaluated two distinct solutions. We chose to
implement and evaluate in this study only the solution reported with
better performance. As in the KNN filter, this solution is also based
on the k-nearest neighbour algorithm. However, instead of filtering
individual examples, this solution filters the most similar projects.
Each project is represented by its distributional characteristics,
given by the mean and standard deviation of each attribute. The authors
suggest a neighbourhood size of $50\%$ from the available projects
to compose the training data.
\item \emph{2012Ma~\cite{Ma2012}:} also known as Transfer Naive Bayes,
this solution introduces the concept of data gravitation to prioritize
and weight training examples. The weights of the training examples
are inversely proportional to their distances from the test set. Given
the indicator $h$:
\[
h(a_{ij})=\begin{cases}
1, & \text{if}\ min_{j}\leq a_{ij}\leq max_{j}\\
0, & \text{otherwise}
\end{cases}
\]
where $a_{ij}$ is the $j$th attribute of the training example $x_{i}$,
and $min_{j}$/$max_{j}$ are the minimum/maximum values of the $j$th
attribute across all test examples. The function $h$ indicates whether
$a_{ij}$ is within the range of values of the test examples. The
distance $s_{i}$ of a training example from the test set is given
by:
\[
s_{i}=\sum_{j=1}^{k}h(a_{ij})
\]
 where $k$ is the number of attributes. Then, the weighting measure
is given by:
\[
w_{i}=\frac{s_{i}}{(k-s_{i}+1)^{2}}
\]
According to this formula, higher weights are given to the more similar
training examples. A weighted Naive Bayes classifier is then constructed
according to the weights given by $w_{i}$.
\end{itemize}
The solutions 2008Watanabe, 2009Turhan, 2009Cruz, 2013He and 2013Herbold
are not dependent on a specific classifier. The solution 2012Ma, however,
is intrinsically associated with the Naive Bayes algorithm. Therefore,
in order to provide a fair comparison, we combined each of the independent
solutions with the five most popular classifiers in the literature
of defect prediction \cite{Malhotra2015}: Random Forest (RF); Support
Vector Machines (SVM); Multilayer Perceptron (MLP); C4.5 (C45); and
Naive Bayes (NB). In Appendix \ref{tab:five-classifiers}, we
briefly describe each classifier. Detailed information can be found
in a specialized machine learning literature \cite{Maimon2010,Sammut2017}.

\begin{figure}[!t]
\centering
\includegraphics[scale=0.37]{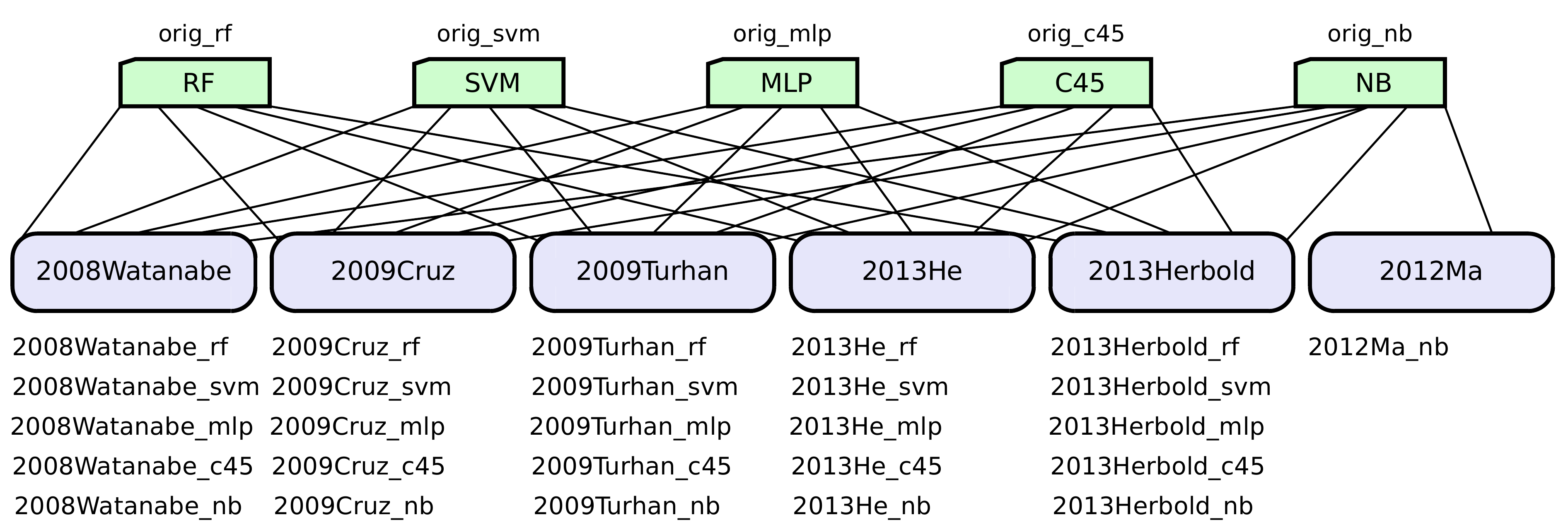}
\caption{\label{fig:combinations}All 31 evaluated CPDP methods derived from
the combination of six transfer learning solutions and five classifiers.}
\end{figure}

For all five classifiers we used the implementation and original parameters
provided in the RWeka Package\footnote{\url{https://CRAN.R-project.org/package=RWeka}}.
As a baseline, we also evaluate the performance of each classifier
in its original form (Orig), with no transfer learning solution. In
total, we evaluate 31 methods, as illustrated in Figure \ref{fig:combinations}.

\subsubsection{Performance Measure\label{sub:Performance-Measure-1}}

We evaluate the predictive performance by means of the AUC (Area Under
the Receiver Operating Characteristic Curve) measure. We chose this
performance measure for three main reasons. First, the AUC measure
provides a single scalar value balancing the influence of True Positive
Rate (TPR) (benefit) and False Positive Rate (FPR) (cost) \cite{Ma2012}.
Further, the use of a single value facilitates the comparison across
models. Second, the AUC measure is not sensitive to the thresholds
used as cutoff parameters. Traditional measures, such as Precision
and Recall, demands a cutoff probability configuration (or prediction
threshold) that defines when an example is classified as defective
or not-defective. Commonly, this parameter is configured to a probability
of 0.5. However, this arbitrary configuration can introduce bias on
the performance analysis \cite{Gunarathna2016}. Lastly, the AUC
measure is also robust to class imbalance, which is frequently present
in software defect prediction datasets \cite{Malhotra2015}. The
datasets used in this study also present class imbalance, as shown
in Table \ref{tab:ap-projs}. An AUC value of $1$ means a perfect
classifier and a $0.5$ means a random classifier.

\subsubsection{Experimental Design\label{sub:Experiment-Design}}

The experimental design is shown in Figure \ref{tab:Pseudocode}.
We aim to evaluate the 31 CPDP methods applied to all the 47 project
versions. For this, we follow a variation of the leave-one-out cross-validation
procedure \cite{dasDores2016}, here called Cross-project Leave-One-Out
(CPLOO). In this variation, the training set contains all the available
project versions except the versions of the same project of the current
test dataset, as described below.

First, we joined all datasets in one unique Cross-Project Dataset
($CPD$). Then, we conducted a many-to-one cross-project analysis
\cite{Gunarathna2016}. Consider a project $\mathcal{P}$, the set
$\mathcal{V_{P}}$ of all versions of $\mathcal{P}$, and a specific
version $v_{i}$ of $\mathcal{V_{P}}$. For each $v_{i}\in\mathcal{V_{P}}$,
$1\leq i\leq|\mathcal{V_{P}}|$, we used $v_{i}$ as test and $CPD\backslash\mathcal{V_{P}}$
( $\mbox{\ensuremath{\mathcal{V_{P}}}}$ subtracted from $CPD$) as
training. In this way, we can analyse the predictive performance disregarding
any bias from the different versions of a same project. This variation
is important considering our experimentation context: to predict defects
in a project based on defect patterns from external projects. This
approach is also named \emph{strict CPDP} in the literature \cite{Herbold2016}.

\begin{figure}[!t]
\centering
\begin{lstlisting}[basicstyle={\scriptsize\ttfamily},breaklines=true,morekeywords={in, FOR, EACH, END},showstringspaces=false,tabsize=2]
projects = [ant,camel,ckjm,forrest,ivy,jedit,log4j,lucene,pbeans,poi,synapse,tomcat,velocity,xalan,xerces] 

FOR EACH project in projects
	versions[project] = getAllVersions(project)
END FOR

CPD = joinAllProjects(versions) //all cross-project data
CPD = preprocess(CPD) //log transformation

eval_methods = getCPDPMethods() //get all 31 evaluated methods

FOR EACH project in projects
	CPTrain = CPD - versions[project] //all datasets except `project' versions
	FOR EACH version in versions[project]
		FOR EACH method in eval_methods
			predictions = applySolution(method,version,CPTrain)    
			AUC = getPerformance(version, predictions)
		END FOR
	END FOR
END FOR
\end{lstlisting}

\caption{\label{tab:Pseudocode}Pseudocode of the experiment design with Cross-project
Leave One Out (CPLOO) cross-validation.}
\end{figure}

\subsubsection{Statistical Test\label{sub:Statistical-Test}}

The statistical tests used for statistical analysis were the Friedman
test followed by Fisher’s LSD test \cite{Pereira2015}. The use of
Friedman test followed by a Post-hoc analysis is recommended for the
comparison of multiple solutions over multiple datasets \cite{Demsar2006}. 

The performances are organized as a table of rankings in which the
rows represent the datasets (project versions), the columns represent
the compared CPDP methods, and each cell is filled with the respective
performance ranking position of a method for a dataset. In this configuration,
tied performances share the average ranking position. For example,
the following AUC performances $[0.79$, $0.79$, $0.77$, $0.76$,
$0.75$, $0.75]$ are ranked as $[1.5$, $1.5$, $3$, $4$, $5.5$,
$5.5]$.

As mentioned in \citet{Larson2006}, the number of significant digits
is justified by the specific research purpose, the measurement accuracy
and sample size. In this statistical test, we rounded the AUC performances
to only 2 significant digits in order to not differentiate very similar
performances.

The Friedman test verifies whether the ranking performances of all
methods are statistically equivalent. When the null hypotheses is
refused, a Post-hoc analysis is applied to define the solutions with
significant difference of performance. Fisher’s LSD test is indicated
as the most powerful method in this scenario according to a recent
study \cite{Pereira2015}. In this test, we considered the confidence
of $95\%$ (i.e., $\text{\emph{p-value}}\leq0.05$).

\subsection{Results\label{sub:cpdp-results}}

\medskip{}

\emph{RQ1: Which CPDP methods perform better across datasets?} 

\medskip{}

Two main factors are compared in this performance analysis. First,
we extract the AUC performance for each of the 31 CPDP methods applied
for all the 47 software projects (or datasets). Then, we extracted
the performance ranking by ordering the AUC performances of the evaluated
methods for each dataset. Table \ref{tab:Performance-obtained-for}
presents the performance mean (and standard deviation) for each CPDP
method. The methods are ordered according to the mean rank.

First, we can observe a generally better performance (across datasets)
of CPDP methods based on the NB learner, including \emph{orig\_nb}.
All 7 methods combined with NB performed among the top 10 positions.
In contrast, the performances based on the C45 classifier performed
among the last positions, except for \emph{2013He\_c45} (position
17). The other classifiers are spread along mixed positions with no
clear assignments. Although the first positions in this table present
close AUC mean values, it is clear the distance between the best and
worst AUC mean performances. The method on the last position performed
close to a random classifier (when $\text{AUC}=0.5$). In this study,
we assume, for practical use, that an AUC $\geq0.75$ represents a
successful model. Although the success of a predictive model is relative
to the application domain, it is commonly set to a performance greater
than 75\% in the CPDP literature \cite{Zimmermann2009,Peters2013,Porto2016b}.

\begin{table}[!t]
\centering
\caption{\label{tab:Performance-obtained-for}Performance Obtained for Each
of the 31 CPDP Methods Considering the Mean (and Standard Deviation)
Values for All 47 Software Projects.}
\begin{tabular}{rlccc}\hline \textbf{Pos.} & \textbf{Method}   & \textbf{Mean Rank} & \textbf{Mean AUC} \\ \hline 1             & 2012Ma\_nb        & 7.57 ($\pm$5.63)   & 0.771 ($\pm$0.10) \\ \hline 2             & 2013He\_rf        & 8.26 ($\pm$6.02)   & 0.766 ($\pm$0.09) \\ \hline 3             & 2009Turhan\_nb    & 9.48 ($\pm$6.12)   & 0.761 ($\pm$0.09) \\ \hline 4             & 2013He\_svm       & 9.90 ($\pm$6.15)   & 0.765 ($\pm$0.09) \\ \hline 5             & 2013Herbold\_nb   & 10.00 ($\pm$5.86)  & 0.760 ($\pm$0.09) \\ \hline 6             & 2009Cruz\_nb      & 10.38 ($\pm$6.03)  & 0.757 ($\pm$0.09) \\ \hline 7             & orig\_nb          & 10.52 ($\pm$5.62)  & 0.758 ($\pm$0.09) \\ \hline 8             & 2013He\_nb        & 11.47 ($\pm$5.88)  & 0.756 ($\pm$0.09) \\ \hline 9             & 2008Watanabe\_nb  & 11.55 ($\pm$6.83)  & 0.752 ($\pm$0.10) \\ \hline 10            & orig\_rf          & 11.74 ($\pm$6.00)  & 0.756 ($\pm$0.09) \\ \hline 11            & 2008Watanabe\_rf  & 12.04 ($\pm$6.68)  & 0.753 ($\pm$0.10) \\ \hline 12            & 2013He\_mlp       & 12.82 ($\pm$8.10)  & 0.753 ($\pm$0.09) \\ \hline 13            & 2013Herbold\_rf   & 12.84 ($\pm$7.10)  & 0.748 ($\pm$0.08) \\ \hline 14            & 2009Turhan\_rf    & 13.03 ($\pm$7.56)  & 0.745 ($\pm$0.10) \\ \hline 15            & orig\_mlp         & 14.86 ($\pm$7.37)  & 0.739 ($\pm$0.09) \\ \hline 16            & 2009Cruz\_rf      & 14.97 ($\pm$5.92)  & 0.737 ($\pm$0.10) \\ \hline 17            & 2013He\_c45       & 15.73 ($\pm$8.27)  & 0.734 ($\pm$0.09) \\ \hline 18            & 2013Herbold\_mlp  & 17.00 ($\pm$6.96)  & 0.729 ($\pm$0.08) \\ \hline 19            & 2009Turhan\_svm   & 17.72 ($\pm$9.38)  & 0.702 ($\pm$0.13) \\ \hline 20            & 2009Cruz\_svm     & 18.17 ($\pm$7.00)  & 0.717 ($\pm$0.10) \\ \hline 21            & orig\_svm         & 18.17 ($\pm$7.00)  & 0.717 ($\pm$0.10) \\ \hline 22            & 2008Watanabe\_mlp & 18.78 ($\pm$6.99)  & 0.714 ($\pm$0.09) \\ \hline 23            & 2008Watanabe\_svm & 19.14 ($\pm$7.05)  & 0.706 ($\pm$0.11) \\ \hline 24            & 2013Herbold\_svm  & 19.26 ($\pm$9.58)  & 0.680 ($\pm$0.13) \\ \hline 25            & 2009Cruz\_mlp     & 21.10 ($\pm$6.92)  & 0.693 ($\pm$0.10) \\ \hline 26            & 2009Turhan\_mlp   & 21.31 ($\pm$8.26)  & 0.688 ($\pm$0.10) \\ \hline 27            & orig\_c45         & 24.87 ($\pm$6.88)  & 0.639 ($\pm$0.09) \\ \hline 28            & 2013Herbold\_c45  & 25.06 ($\pm$6.99)  & 0.636 ($\pm$0.10) \\ \hline 29            & 2009Turhan\_c45   & 25.16 ($\pm$7.22)  & 0.633 ($\pm$0.10) \\ \hline 30            & 2009Cruz\_c45     & 25.69 ($\pm$6.82)  & 0.624 ($\pm$0.09) \\ \hline 31            & 2008Watanabe\_c45 & 27.39 ($\pm$4.77)  & 0.602 ($\pm$0.08) \\ \hline \end{tabular}
\end{table}

\begin{figure}[!t]
\centering
\includegraphics[scale=0.45]{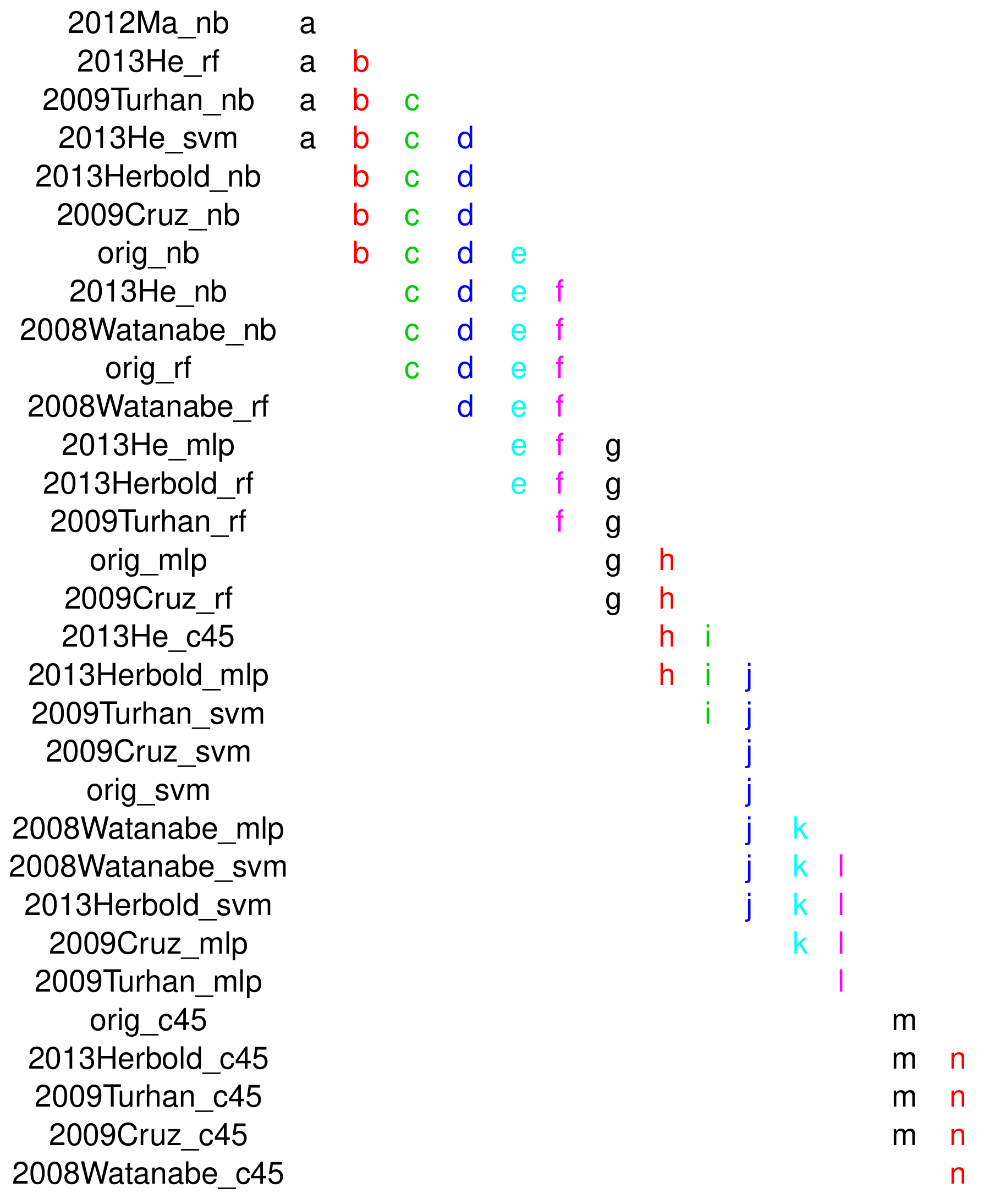}
\caption{\label{fig:Pairwise-Post-hoc-Fisher=002019s}Pairwise Post-hoc Fisher’s
LSD test. Each letter represents a group of methods with no statistically
significant difference.}
\end{figure}

As discussed in Section \ref{sub:Statistical-Test}, we analysed the
statistical significance of results based on the Friedman test. The
null hypothesis assumes all performances as equivalent. The alternative
hypothesis is that at least one pair of predictive models has different
performance. The null hypothesis is rejected with a \emph{p-value}$<$2.2e-16.
Therefore, we analyse the pairwise difference of performances with
the Fisher’s LSD test. The result is presented in Figure \ref{fig:Pairwise-Post-hoc-Fisher=002019s}.
The alphabet letters group the methods with no significant difference. 

No significant difference was found between\emph{ 2012Ma\_nb} and
the next 3 methods although it presents better performance in relation
to all remaining methods - including \emph{orig\_nb}. This differentiation
in relation to the original learner is important since it elucidates
the real gain produced by each transfer learning solution. For example,
except for \emph{2012Ma\_nb}, no significant difference between the
CPDP methods using NB and \emph{orig\_nb} was found - as stated in
group 'c'. 

This relation between transfer learning solutions and their respective
learner baseline is better exposed in Table \ref{tab:cpdp_vs_baseline}.
When there is no significant difference, the relation is represented
by the symbol ``/''. Otherwise, it is filled with ``(+)'' or ``($-$)'',
meaning a better or worse mean rank position, respectively.

\begin{table}[!t]
\centering
\caption{\label{tab:cpdp_vs_baseline}Comparison of CPDP Methods in Relation
to the Baseline Original Learners, Based on the Statistical Fisher’s
LSD test. The Symbol ``/'' Means no Statistical Difference. The
Symbols ``(+)'' or ``($-$)'' Mean Statistical Difference with
Better or Worse Mean Rank, Respectively.}
\begin{tabular}{lccccc}  \hline         & \textbf{rf} & \textbf{svm} & \textbf{mlp} & \textbf{c45} & \textbf{nb} \\ \hline orig         & ---         & ---          & \textbf{---}          & ---          & ---         \\ \hline 2008Watanabe & /           & /            & ($-$)          & ($-$)          & /           \\ \hline 2009Cruz     & ($-$)         & /            & ($-$)          & /            & /           \\ \hline 2009Turhan   & /           & /            & ($-$)          & /            & /           \\ \hline 
2013He       & \textbf{(+)}         & \textbf{(+)}          & /            & \textbf{(+)}          & /           \\ \hline 2013Herbold  & /           & /            & /            & /            & /           \\ \hline 2012Ma       & ---         & ---          & ---          & ---          & \textbf{(+)}         \\ \hline \end{tabular}
\end{table}

From this table, we can observe that most of the applied transfer
learning solutions do not lead to significant difference of performance
in relation to the original classifiers. Moreover, some solutions
diminish the performances when associated to a learner, as observed
for the MLP classifier with the solutions \emph{2008Watanabe}, \emph{2009Cruz},
and \emph{2009Turhan}. On the other hand, the solution \emph{2013He}
significantly improves the performances of the classifiers RF, SVM,
and C45. The solution \emph{2012Ma} also improves the performance
when associated to the NB classifier. 

Based upon the statistical analysis presented in Figure \ref{fig:Pairwise-Post-hoc-Fisher=002019s},
the methods \emph{2012Ma\_nb}, \emph{2013He\_rf}, \emph{2009Turhan\_nb},
and \emph{2013He\_svm} presented the best performances across datasets
although the method \emph{2009Turhan\_nb} did not present significant
difference in relation to its respective original learner baseline.

However, other criteria can be considered for a comparative analysis.
For example, in their original published work \cite{Ma2012,He2013},
both\emph{ 2012Ma} and \emph{2013He} presented lower computation time
cost in relation to the \emph{2009Turhan} solution, although \emph{2013He}
presented higher complexity in relation to \emph{2012Ma}. On the other
hand, \emph{2013He} is more robust to redundant and irrelevant attributes
in relation to \emph{2012Ma}. First, because \emph{2013He} has an
internal procedure to filter the most relevant attributes. In addition,
both classifiers RF and SVM are known to be robust in this context.
Second, because \emph{2012Ma} is sensitive to redundant and irrelevant
attributes in two points: its internal weighting procedure, based
on the attributes relation between testing and training examples;
and its assumption of independence between attributes inherited from
the Naive Bayes algorithm.

\medskip{}

\emph{RQ2: Do the best CPDP methods perform better for the same datasets? }

\medskip{}

To answer this question, we use of the information shown in Figure
\ref{fig:Pairwise-Post-hoc-Fisher=002019s}. We evaluate the performances
of the CPDP methods: \emph{2012Ma\_nb}, \emph{2013He\_rf}, \emph{2009Turhan\_nb},
and \emph{2013He\_svm}; referring to the four methods with better
ranking across datasets. Then, for each dataset, we associate the
best AUC performance obtained among these four CPDP methods. This
association of best methods for each dataset is presented in Table
\ref{tab:best_methods}. As already mentioned in Section \ref{sub:Statistical-Test},
the AUC performance comparison is based on 2 significant digits only.
In this way, we do not differentiate very similar performances.

\begin{table}[!t]
\centering
\caption{\label{tab:best_methods}Best AUC Balue and Respective Best CPDP Methods
by Project, Considering the Four Methods with Better Ranking Across
Datasets. The AUC Values were Rounded for Only 2 Significant Digits
in Order to not Differentiate Similar Performances. }
\begin{tabular}{l cc L{2.9cm}} \hline \textbf{Project}      & \textbf{Best AUC} & \textbf{\# Best Methods} & \textbf{Best Methods}                               \\ \hline ant-1.3.csv           & \textbf{0.88}     & 1                        & 2012Ma\_nb                                          \\ \hline ant-1.4.csv           & 0.68              & 1                        & 2012Ma\_nb                                          \\ \hline ant-1.5.csv           & \textbf{0.84}     & 1                        & 2012Ma\_nb                                          \\ \hline ant-1.6.csv           & \textbf{0.86}     & 1                        & 2012Ma\_nb                                          \\ \hline ant-1.7.csv           & \textbf{0.85}     & 1                        & 2012Ma\_nb                                          \\ \hline camel-1.0.csv         & \textbf{0.84}     & 1                        & 2009Turhan\_nb                                      \\ \hline camel-1.2.csv         & 0.64              & 1                        & 2013He\_rf                                          \\ \hline camel-1.4.csv         & 0.74              & 1                        & 2009Turhan\_nb                                      \\ \hline camel-1.6.csv         & 0.68              & 1                        & 2013He\_rf                                          \\ \hline ckjm-1.8.csv          & \textbf{0.96}     & 1                        & 2013He\_svm                                         \\ \hline forrest-0.7.csv       & \textbf{0.81}     & 1                        & 2009Turhan\_nb                                      \\ \hline forrest-0.8.csv       & \textbf{0.91}     & 1                        & 2013He\_svm                                         \\ \hline ivy-1.1.csv           & \textbf{0.81}     & 2                        & 2013He\_rf, 2012Ma\_nb                              \\ \hline ivy-1.4.csv           & \textbf{0.8}      & 1                        & 2013He\_rf                                          \\ \hline ivy-2.0.csv           & \textbf{0.85}     & 2                        & 2012Ma\_nb, 2013He\_rf                              \\ \hline jedit-3.2.1.csv       & \textbf{0.89}     & 1                        & 2012Ma\_nb                                          \\ \hline jedit-4.0.csv         & \textbf{0.87}     & 2                        & 2013He\_rf, 2009Turhan\_nb                          \\ \hline jedit-4.1.csv         & \textbf{0.91}     & 2                        & 2013He\_rf, 2013He\_svm                             \\ \hline jedit-4.2.csv         & \textbf{0.92}     & 4                        & 2013He\_rf, 2012Ma\_nb, 2009Turhan\_nb, 2013He\_svm \\ \hline jedit-4.3.csv         & \textbf{0.9}      & 1                        & 2013He\_svm                                         \\ \hline log4j-1.0.csv         & \textbf{0.89}     & 1                        & 2009Turhan\_nb                                      \\ \hline log4j-1.1.csv         & \textbf{0.86}     & 1                        & 2012Ma\_nb                                          \\ \hline log4j-1.2.csv         & \textbf{0.88}     & 1                        & 2012Ma\_nb                                          \\ \hline lucene-2.0.csv        & \textbf{0.81}     & 1                        & 2012Ma\_nb                                          \\ \hline lucene-2.2.csv        & \textbf{0.75}     & 2                        & 2013He\_svm, 2013He\_rf                             \\ \hline lucene-2.4.csv        & \textbf{0.8}      & 2                        & 2013He\_rf, 2013He\_svm                             \\ \hline pbeans-1.0.csv        & 0.68              & 1                        & 2009Turhan\_nb                                      \\ \hline pbeans-2.0.csv        & \textbf{0.81}     & 1                        & 2009Turhan\_nb                                      \\ \hline poi-1.5.csv           & \textbf{0.76}     & 1                        & 2012Ma\_nb                                          \\ \hline poi-2.0RC1.csv        & 0.7               & 2                        & 2013He\_svm, 2013He\_rf                             \\ \hline poi-2.5.1.csv         & \textbf{0.79}     & 1                        & 2012Ma\_nb                                          \\ \hline poi-3.0.csv           & \textbf{0.85}     & 3                        & 2013He\_rf, 2013He\_svm, 2012Ma\_nb                 \\ \hline synapse-1.0.csv       & \textbf{0.79}     & 1                        & 2012Ma\_nb                                          \\ \hline synapse-1.1.csv       & 0.68              & 3                        & 2012Ma\_nb, 2013He\_svm, 2013He\_rf                 \\ \hline synapse-1.2.csv       & \textbf{0.77}     & 1                        & 2009Turhan\_nb                                      \\ \hline tomcat-6.0.389418.csv & \textbf{0.82}     & 1                        & 2012Ma\_nb                                          \\ \hline velocity-1.4.csv      & 0.62              & 2                        & 2013He\_svm, 2013He\_rf                             \\ \hline velocity-1.5.csv      & 0.71              & 2                        & 2012Ma\_nb, 2009Turhan\_nb                          \\ \hline velocity-1.6.1.csv    & \textbf{0.76}     & 1                        & 2013He\_svm                                         \\ \hline xalan-2.4.0.csv       & \textbf{0.81}     & 1                        & 2013He\_rf                                          \\ \hline xalan-2.5.0.csv       & 0.68              & 1                        & 2013He\_rf                                          \\ \hline xalan-2.6.0.csv       & \textbf{0.76}     & 1                        & 2013He\_rf                                          \\ \hline xalan-2.7.0.csv       & \textbf{0.85}     & 1                        & 2013He\_rf                                          \\ \hline xerces-1.2.0.csv      & 0.53              & 1                        & 2013He\_rf                                          \\ \hline xerces-1.3.0.csv      & 0.72              & 1                        & 2009Turhan\_nb                                      \\ \hline xerces-1.4.4.csv      & \textbf{0.76}     & 2                        & 2009Turhan\_nb, 2013He\_svm                         \\ \hline xerces-init.csv       & 0.68              & 1                        & 2013He\_rf                                          \\ \hline Mean                  & 0.791             & 1.36                     & -                                                   \\ \hline \end{tabular}
\end{table}

\begin{figure}[!t]
\centering
\includegraphics[scale=0.45]{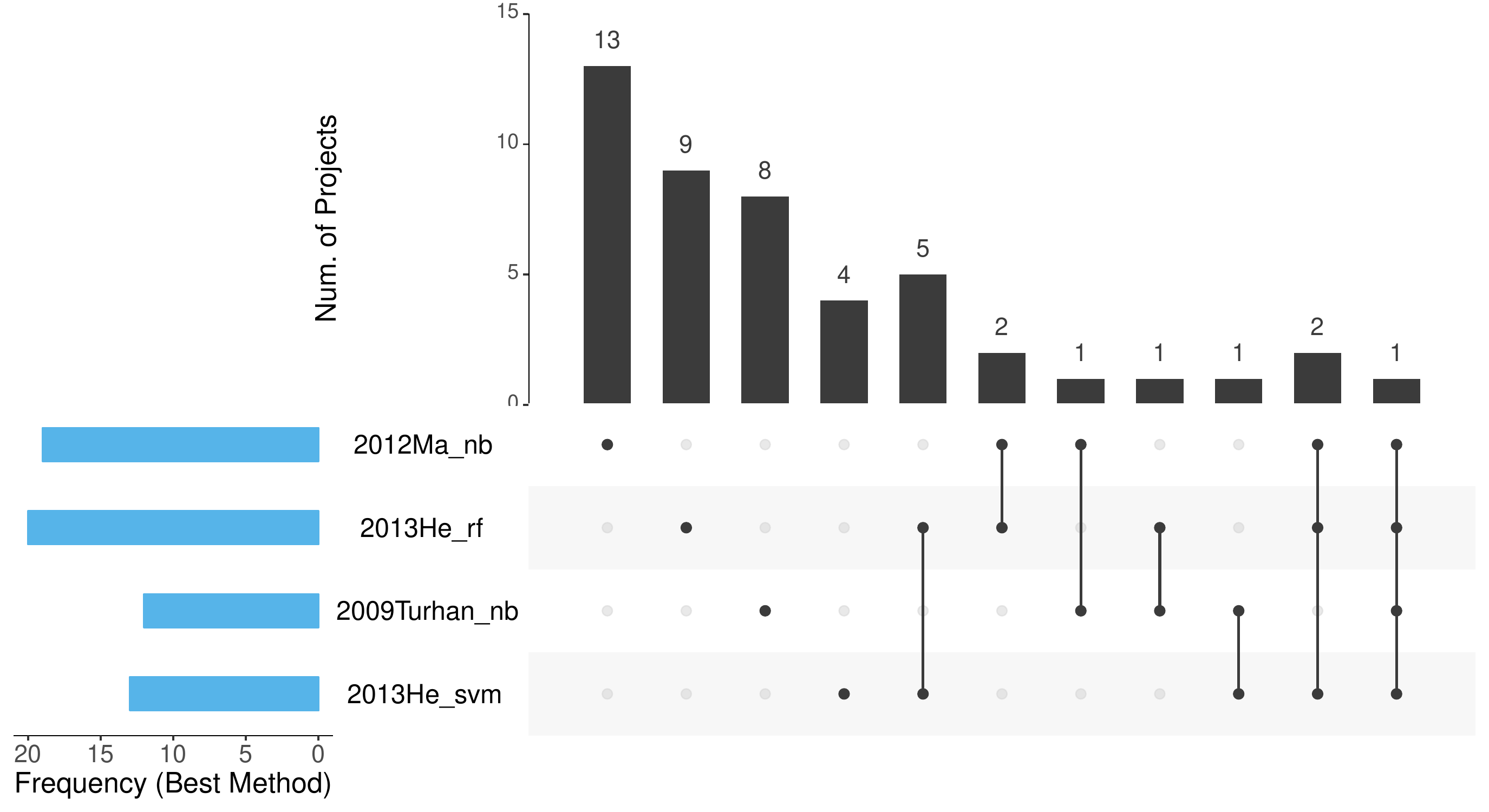}
\caption{\label{fig:intersection_sets}Data of Table \ref{tab:best_methods}
organized by intersection sets. Each CPDP method represents a set
of all projects in which it achieved the best performance. The bars
represent the distribution of all 47 projects along the four methods.
When a project has more than one best method, the intersection is
represented by the connected dots.}
\end{figure}

We can extract some information from this table. First, the best CPDP
method for a project version is not necessarily the same for all versions.
More than one CPDP method can achieve the best AUC performance for
the same project version. From all the 47 project versions: $72\%$
achieved a successful performance with $\text{AUC}\geq0.75$; $21\%$
performed in the range $0.65\leq\text{AUC}<0.75$; and only $6\%$
performed with AUC below $0.65$. Last but not least, there is no
general better performance. This statement is better visualized in
Figure \ref{fig:intersection_sets}. In this figure, each CPDP method
represents a set of all projects in which it achieved the best performance.
The intersections between sets (i.e when a project version has more
than one best method) are represented by the connected dots.

From all the 47 project versions: 28\% achieved the best performance
exclusively with \emph{2012Ma\_nb}; 19\% with \emph{2013He\_rf}; 17\%
with \emph{2009Turhan\_nb}; 9\% with \emph{2013He\_svm}; and 27\%
share the best performance with more than one CPDP method. As we can
see, no method always performs best. 

In the next section we investigate to what extent a meta-learning
solution could predict the best CPDP method according to the project
characteristics. The next experiments are based on the results discussed
in this section.

\section{Meta-learning for CPDP \label{sec:Meta-learning-for-CPDP}}

In this section we propose a meta-learning architecture for the recommendation
of CPDP methods. First, we present the general methodology and details
related to the construction of the meta-model. Next, we evaluate the
performance of the proposed solution. The experimental design is presented
in Section \ref{sub:Experimental-Setup} and the results are discussed
in Section \ref{sub:Results}.

\subsection{Proposed Methodology \label{sub:Proposed-Architecture}}

As mentioned in Section \ref{sub:Meta-learning}, meta-learning is
commonly applied for the recommendation of base learners in specific
tasks. In this study, we propose a different application of meta-learning,
designed for the recommendation of CPDP methods. The process is based
on the general meta-learning approach proposed in \citet{Kalousis2002}.
Three main differences can be highlighted: 1) the performances are
obtained from external project datasets with CPLOO cross-validation
instead of a cross-validation within the datasets (see Section \ref{sub:Experiment-Design});
2) we adopted unsupervised meta-attributes instead of the traditional
supervised meta-attributes, commonly used in the literature \cite{dasDores2016}
(see Section \ref{sub:Meta-features}); and 3) the meta-target characterizes
a multi-label classification task \cite{Tsoumakas2007} (see Section
\ref{sub:Meta-target}). 

\begin{figure}[!t]
\centering
\includegraphics[scale=0.45]{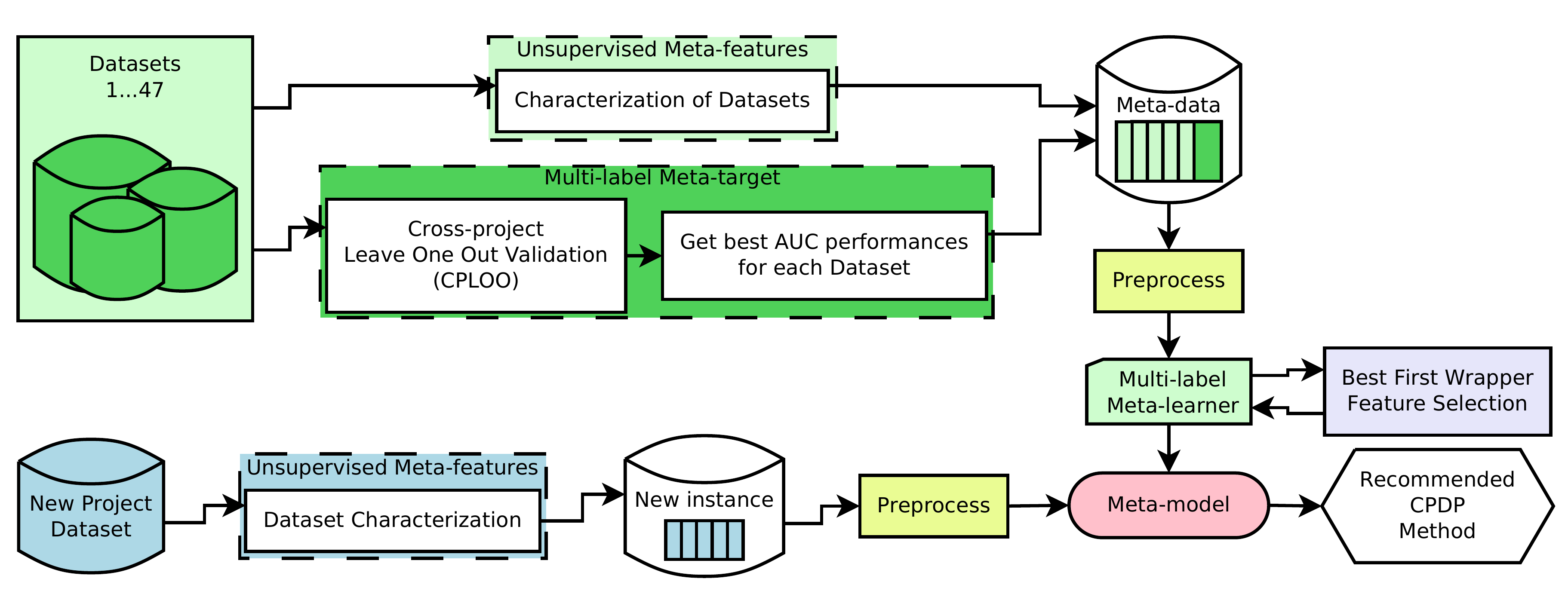}
\caption{\label{fig:Meta-learning-proposal}Proposed architecture: meta-learning
for CPDP.}
\end{figure}

Figure \ref{fig:Meta-learning-proposal} presents the general architecture
of the proposed solution. The meta-target is determined by the performances
obtained from the input datasets. The meta-attributes are also extracted
from the characterization of these datasets. Next to the meta-data
preprocessing, a multi-label meta-learner is applied. This meta-learner
is associated to a wrapper attribute selection in order to select
a subset of meta-attributes to compose the final meta-model. The meta-model
can be then used to recommend suitable CPDP methods for new project
datasets. Once the meta-model is constructed, only the new project
datasets need to be characterized and preprocessed before predicting
a suitable CPDP method. Greater details and related issues are discussed
in the following subsections.

\subsubsection{Input Datasets\label{sub:Meta-data}}

The collection of datasets (meta-examples) is composed of the 47 projects
already presented in Section \ref{sub:Software-Projects}. The amount
of meta-examples is important since the more available information
more effective and generalizable can be the meta-model. However, as
stated in \citet{Sammut2017}, scarce training data for the meta-learning
task is a common aspect in this context. Although there is no recommended
minimum amount of data, more than 50 datasets are desirable for a
meta-learning analysis \cite{dasDores2016}. Thereby, the available
amount of data is a limitation for this study.

\subsubsection{Meta-attributes\label{sub:Meta-features}}

As already mentioned in Section \ref{sub:Meta-learning}, in this
study we focus on the unsupervised characterization of datasets. Contrary
to the supervised characterization, the unsupervised approach does
not demand the previous information about the class attribute (defective
or not-defective). This approach is important in the context of this
study since we work with the assumption that historical defect information
may not be available for a software company. This approach, however,
is little explored in the meta-learning literature \cite{Santos2016}.

Within the context of this study, we evaluate two different sets of
meta-attributes. Both are related to continuous data and can be automatically
extracted from the original dataset. 

The first meta-attributes set (here called MS-Dist) corresponds to
the direct application of distributional measures over all the dataset
attributes. For this, we consider 5 distributional measures: mean
(mean), standard deviation (sd), median (med), maximum (max) and minimum
(min). This approach leads to 100 meta-attributes, considering all
5 distributional measures applied for all the 20 attributes of a dataset.
We also included the size (number of examples) of a dataset, totalizing
101 meta-attributes. The use of distributional measures to characterize
datasets was already proposed in the CPDP literature but for different
purposes \cite{He2012,He2013,Herbold2013}.

The second meta-attributes set (here called MS-Uns) is proposed in
\citet{Santos2016} and covers different characteristics of a dataset
including general measures, statistical measures and clustering based
measures. Originally, this set includes 53 unsupervised meta-attributes
applied in the context of active learning. From the original set,
we selected 44 measures applicable to the context of this study. The
selected subset of meta-attributes is presented in Table \ref{tab:mf-uns}.
The measures matching the pattern $M_{\Omega}$ (e.g., $\text{sd}_{max}$,
$\text{corr}_{min/max}$, $\text{skew}_{mean}$) are obtained in two
steps. First, the measure $M$ is extracted for each element of the
dataset, generating a vector of values (e.g., the standard deviation
extracted for all the attributes or the correlation extracted for
all the pairs of attributes). Then, a distributional measure (e.g.,
min, max) is applied over this vector of values, generating a single
value. The numbers $1$, $1.5$, and $2$ attached to the measures
kurt, conn, dunn, and silh represents an internal parametrization
indicating the proportion of clusters per class.

\begin{table}[!t]
\centering
\caption{\label{tab:mf-uns} Unsupervised Meta-attributes Set (MS-Uns), Where
$\mathcal{U}$ Represents a Dataset with $n$ Examples, $x\in\mathcal{U}$
Represents an Example of $\mathcal{U}$, and $1<j<20$ Represents
an Attribute of $\mathcal{U}$. (Adapted from \cite{Santos2016})}

\begin{tabular}{L{4cm}L{4.5cm}C{7cm}} \hline  \textbf{Meta-feature} & \textbf{Name} & \textbf{Formula}\tabularnewline \hline  {$\text{size}$} & {Size (number of instances)} & {$n=|\mathcal{{U}}|$}\tabularnewline \hline  {$\text{size}_{lg}$} & {Logarithm of size} & {$log\,n$}\tabularnewline \hline  {$\text{mean}_{min}$, $\text{mean}_{max}$, $\text{mean}_{mean}$, $\text{mean}_{min/max}$} & {Mean } & {$\mu_{j}=\frac{1}{n}\sum\limits _{x\in\mathcal{U}}x_{j}$}\tabularnewline \hline  {$\text{sd}_{min}$, $\text{sd}_{max}$, $\text{sd}_{mean}$, $\text{sd}_{min/max}$} & {Standard deviation } & {$\sigma_{j}=\frac{1}{n}\sum\limits _{x\in\mathcal{U}}(x_{j}-\mu_{j})^{2}$}\tabularnewline \hline  {$\text{entr}_{min}$, $\text{entr}_{max}$, $\text{entr}_{mean}$, $\text{entr}_{min/max}$} & {Normalized entropy} & {$\text{entr}_{j}=\frac{-1}{log\,n}\sum\limits _{x\in\mathcal{U}}x_{j}logx_{j}$}\tabularnewline \hline  {$\text{corr}_{min}$, $\text{corr}_{max}$, $\text{corr}_{mean}$, $\text{corr}_{min/max}$} & {Correlation between features } & {$\text{corr}_{jk}=(\sigma_{j}^{2}.\sigma_{k}^{2})^{-\frac{1}{2}}\sum\limits _{x\in\mathcal{U}}(x_{j}-\mu_{j})(x_{k}-\mu_{k})$}\tabularnewline \hline  {$\text{skew}_{min}$, $\text{skew}_{max}$, $\text{skew}_{mean}$, $\text{skew}_{min/max}$} & {Skewness} & {$\text{skew}_{j}=\frac{n}{(n-1)(n-2)}\sum\limits _{x\in\mathcal{U}}\frac{(x_{j}-\mu_{j})^{3}}{\sigma_{j}^{3}}$}\tabularnewline \hline  {$\text{kurt}_{min}$, $\text{kurt}_{max}$, $\text{kurt}_{mean}$, $\text{kurt}_{min/max}$} & {Kurtosis} & {$\text{kurt}_{j}=\left\{ \frac{n(n+1)}{(n-1)(n-2)(n-3)}\sum\limits _{x\in\mathcal{U}}\frac{(x_{j}-\mu_{j})^{4}}{\sigma_{j}^{4}}\right\} -\frac{3(n-1)^{2}}{(n-2)(n-3)}$}\tabularnewline \hline  {$\text{conn}_{k1}$, $\text{conn}_{k1.5}$, $\text{conn}_{k2}$} & {Connectivity }\emph{k-means} & {Cluster validity measure \cite{Xu2009}}\tabularnewline \hline  {$\text{conn}_{h1}$, $\text{conn}_{h1.5}$, $\text{conn}_{h2}$} & {Connectivity }\emph{hierarc. clust.} & {Cluster validity measure \cite{Xu2009}}\tabularnewline \hline  {$\text{dunn}_{k1}$, $\text{dunn}_{k1.5}$, $\text{dunn}_{k2}$} & {Dunn index }\emph{k-means} & {Cluster validity measure \cite{Dunn1974}}\tabularnewline \hline  {$\text{dunn}_{h1}$, $\text{dunn}_{h1.5}$, $\text{dunn}_{h2}$} & {Dunn index}\emph{ hierarc. clust.} & {Cluster validity measure \cite{Dunn1974}}\tabularnewline \hline  {$\text{silh}_{k1}$, $\text{silh}_{k1.5}$, $\text{silh}_{k2}$} & {Silhouette }\emph{k-means} & {Cluster validity measure \cite{Rousseeuw1987}}\tabularnewline \hline  {$\text{silh}_{h1}$, $\text{silh}_{h1.5}$, $\text{silh}_{h2}$} & {Silhouette}\emph{ hierarc. clust.} & {Cluster validity measure \cite{Rousseeuw1987}}\tabularnewline \hline  \end{tabular}
\end{table}

\subsubsection{Meta-target\label{sub:Meta-target}}

The meta-target is obtained from the experiment results presented
in Section \ref{sec:CPDP-Methods-Performance}. From the 31 evaluated
CPDP methods, we consider four possible labels for a project: \emph{2012Ma\_nb},
\emph{2013He\_rf}, \emph{2009Turhan\_nb}, and \emph{2013He\_svm}.
These four methods refer to the best-ranked methods across datasets,
as discussed in Section \ref{sub:cpdp-results}. 

Another important characteristic in this study relates to the meta-target
designed for a multi-label scenario. As mentioned in Section \ref{sub:Statistical-Test},
we rounded the AUC performances for only 2 significant digits. On
the one hand, this leads to a more accurate analysis since we do not
differentiate CPDP methods with very similar performance. On the other
hand, however, this leads to a multi-label classification task since
more than one label can be associated to the same meta-example (see
Section \ref{sub:Multi-label-Classification}). Table \ref{tab:best_methods}
presents the best AUC performance obtained for each project and the
respective best methods with equivalent performances. 

For this multi-label classification task we applied the Binary Relevance
(BR) transformation method \cite{Luaces2012}. As discussed in Section
\ref{sub:Multi-label-Classification}, this method creates $c$ distinct
datasets ($c=|\mathcal{L}|$, total number of labels), each for one
of the four possible labels. The multi-label problem is then transformed
in four binary classification problems. For each dataset $D_{\lambda_{j}}$,
$1\leq j\leq c$, the class attribute is positive for meta-examples
that belongs to the label $\lambda_{j}$ and negative otherwise.

\subsubsection{Meta-data Preprocessing}

In the preprocessing step, we address two issues. First, we mitigate
the likely influence of the different ranges and scales of data over
the meta-model performance. For this, we applied the z-score normalization
\cite{Nam2013}. In this normalization technique, each meta-attribute
column is centred by subtracting the mean; and also scaled by dividing
each value by the standard deviation. Second, we address the class-imbalance
issue, possibly resulting from the BR transformation method \cite{Zhang2014b}.
For example, consider the label \emph{2009Turhan\_nb}. This method
performed the best AUC for only 26\% of all project datasets (see
Table \ref{tab:best_methods}), which leads to an imbalanced binary
dataset. In order to mitigate this issue, we applied the oversampling
technique \cite{Galar2012}. This technique consists of randomly
duplicating examples of the minority class until the desired class
distribution is achieved. It allows to adjust the class distribution
of a dataset without discarding information. This characteristic is
important in the context of this study considering the limited amount
of meta-examples. On the other hand, the duplication of examples may
lead to overfitting on data.

\subsubsection{Meta-learner\label{sub:Meta-model}}

Given a new software project dataset, represented by the meta-example
$\mathbf{x}_{\mathbf{i}}\in\mathbb{R}^{m\times1}$, the meta-model
must be able to recommend an appropriate CPDP method (or label $\lambda\in\mathcal{L}$
for $\mathbf{x}_{\mathbf{i}}$). For each dataset $D_{\lambda_{j}}$,
$1\leq j\leq c$, generated with BR transformation method (see Section
\ref{sub:Meta-target}), we apply a binary classifier able to generate
the confidence of relevance $C_{ij}$ (the confidence of $\lambda_{j}$
be a relevant label for $\mathbf{x_{i}}$). The final recommended
label $\lambda$ refers to the label $\lambda_{j}$ with higher confidence
of relevance $C_{ij}$.

For this task, we used the Random Forest algorithm \cite{Breiman2001}.
Based on our experiments, this classifier presented the best learning
capacity compared to each of the five classifiers investigated in
this study. This algorithm is also a common choice in the contexts
of software defect prediction \cite{Malhotra2015} and algorithm
recommendation \cite{dasDores2016,Santos2016}.

\subsubsection{Performance Measure\label{sub:Performance-Measure}}

Several performance measures focused on different aspects of multi-label
learning have been proposed in the literature \cite{Wu2016}. In
this study we are specifically interested in evaluating the frequency
(or accuracy) in which the top-ranked (higher confidence of relevance
$C_{ij}$) label is actually among the relevant labels of an example.
This aspect can be obtained from the one-error measure \cite{Schapire2000}.
The one-error measures how many times the predicted label was \emph{not
}in $L_{i}$ (set of relevant labels for $\mathbf{x}_{\mathbf{i}}$).
It is defined as follows:

\[
\text{one-error}(F)=\frac{1}{n}\sum\limits _{i=1}^{n}\llbracket\text{arg max }F(x_{i})\notin L_{i}\rrbracket.
\]

Note that, for single-label classification tasks, the one-error is
identical to the ordinary error measure. Our general goal is to maximize
the accuracy given by $\text{acc}(F)=1-\text{one-error}(F)$.

\subsubsection{Attribute Selection\label{sub:Feature-Selection}}

As mentioned in Section \ref{sub:Meta-data}, the amount of meta-data
available is limited. High dimensional datasets, with redundant and
irrelevant attributes, can lead to an ineffective performance \cite{Parmezan2017}.
Therefore, we also apply an Attribute Selection method over the meta-data
in order to achieve a suitable subset of meta-attributes. 

We apply a Best-First Forward Wrapper strategy, adapted from \citet{Kalousis2001}.
This strategy applies an extensive and systematic search in the state
space of all possible attribute subsets using the Best-First heuristic
\cite{Kohavi1995}. The searching is guided by the estimated accuracy
of each subset, provided by an induction algorithm. To estimate the
accuracy of a given subset we apply the CPLOO cross-validation; where
each version (or meta-example) is tested over a training set containing
all the remaining project versions, excepting the versions of same
project (see Section \ref{sub:Experiment-Design}). The accuracy measure
is presented in Section \ref{sub:Performance-Measure} and the induction
algorithm is discussed in Section \ref{sub:Meta-model}.

At the end, the subset of meta-attributes with highest accuracy is
selected.

\subsection{Performance Evaluation\label{sub:Experimental-Setup}}

In an ideal scenario, we would be able to construct a meta-model based
on the 47 available datasets (as proposed in Section \ref{sub:Proposed-Architecture})
and evaluate its performance with a different set of datasets. However,
this scenario is not possible since this different set of datasets
is not available. In order to approximate the ideal scenario, we evaluate
the proposed solution based on a variation of the CPLOO procedure,
here called meta-CPLOO. We adapted this leave-one-out procedure to
the context of CPDP although it is a common approach in the meta-learning
literature \cite{dasDores2016}.

We separate one project (and its respective versions) for testing
and construct the meta-model based on the remaining project versions.
The meta-model is constructed following the configuration presented
in the previous section. Finally, each version is tested separately
with the respective constructed meta-model. It is important to observe
that $15$ different meta-models (one for each project) will be constructed
to test all the $47$ project versions. In the end, a CPDP method
is recommended for each version.

On the one hand, the meta-CPLOO procedure enables us to estimate the
performance for the proposed solution. On the other hand, it reduces
the already limited amount of meta-data. For example, if we separate
for test the five versions of the Ant project, the meta-model will
be constructed based only on 42 versions. Consequently, the amount
of available data for the internal CPLOO procedures are reduced.

Based on the meta-CPLOO procedure, we evaluate the performance of
the meta-learning solution in two different levels: the meta-level
and the base-level. In the meta-level we evaluate the meta-learning
capacity, i.e., whether the meta-learner can learn from the meta-data
in relation to the defined baselines. In the base-level we evaluate
the performance of the meta-learning solution across datasets in terms
of AUC and compare it in relation to the four considered base CPDP
methods. In each level, we evaluate two different configurations of
meta-learning: MS-Dist and MS-Uns; referring to the two meta-attributes
sets presented in Section \ref{sub:Meta-features}.

\subsubsection{Meta-level\label{sub:Meta-level}}

In the meta-level analysis, we evaluate two sources of results. The
first source is obtained from the attribute selection step (see Section
\ref{sub:Feature-Selection}). For each of the 15 projects, the subset
of attributes with highest accuracy estimate is selected to compose
the meta-model. We compare these accuracy estimates in relation
to a baseline, defined as the majority class. In this context, the
majority class is given by the most frequent label of the respective
meta-data. The statistical significance is verified with the non-parametric
Wilcoxon signed rank test ($\text{\emph{p-value}}\leq0.05$) \cite{Keung2013}.

The second source is obtained from the final recommendations provided
by the proposed solution. We compare the accuracy (i.e., the rate
of correct predictions) obtained with the meta-learning solution in
relation to the majority class. In this context, the majority class
is given by the most frequent label presented in Table \ref{tab:best_methods}.

\subsubsection{Base-level}

In the base-level analysis, we compare the general performance (in
terms of AUC) of the meta-learning solution in relation to the four
considered CPDP methods applied individually. In addition, we compare
all the performances in relation to a random baseline. For each project
version, one of the four evaluated methods is randomly selected. This
process is repeated 30 times. The random baseline is given by the
mean AUC of the respective selected methods. The statistical analysis
is based on the Friedman test followed by Fisher’s LSD test \cite{Pereira2015},
as presented in Section \ref{sub:Statistical-Test}.

\subsection{Results\label{sub:Results}}

\medskip{}

\emph{RQ3: To what extent can meta-learning help us to select the
most suitable CPDP method for a given dataset?}

\medskip{}

\begin{itemize}
\item \emph{RQ3.1: Does the meta-learner learn? (Meta-level)}
\end{itemize}
\medskip{}

To answer this question, we evaluate two sources of results, as discussed
in Section \ref{sub:Meta-level}. First, we compare the accuracies
estimates produced in the attribute selection step (see Section \ref{sub:Feature-Selection})
for each of the 15 generated meta-models. We use these accuracies
to evaluate the learning capacity of the two meta-learning configurations:
\emph{MS-Dist} and \emph{MS-Uns}. We compare their performances in
relation to the majority class baseline. The obtained accuracies are
presented in Table \ref{tab:bf-accuracies}.

\begin{table}[!t]
\centering
\caption{\label{tab:bf-accuracies}The Estimated Accuracies and Selected Subsets
for Each Meta-Model. The Majority Class is Used as a Comparative Reference.}

\begin{tabular}{lcccll} \hline \textbf{}        & \textbf{}         & \multicolumn{2}{c}{\textbf{Best-First - Acc}} & \multicolumn{2}{c}{\textbf{Best-First - Selected Subset}}                                                                         \\ \hline \textbf{Project} & \textbf{Majority} & \textbf{MS-Dist}       & \textbf{MS-Uns}      & \textbf{MS-Dist}                                                 & \textbf{MS-Uns}                                    \\ \hline
ant      & 0.500         & 0.595         & 0.595         &
$\text{dam}_{med}$, $\text{mfa}_{med}$                                   &
$\text{mean}_{min/max}$                                             \\ \hline 
camel    & 0.465         & 0.628         & 0.628         &
$\text{avg\_cc}_{sd}$, $\text{cam}_{med}$                                   & 
$\text{corr}_{mean}$, $\text{kurt}_{mean}$, $\text{silh}_{h1}$                  \\ \hline
ckjm     & 0.435         & 0.609         & 0.739         & $\text{max\_cc}_{max}$, $\text{avg\_cc}_{sd}$                                  & $\text{sd}_{min/max}$, $\text{corr}_{mean}$, $\text{dunn}_{k1.5}$, $\text{silh}_{k1.5}$ \\ \hline forrest  & 0.533         & 0.733         & 0.733         & $\text{npm}_{sd}$, $\text{rfc}_{med}$, $\text{ce}_{med}$                     & $\text{corr}_{mean}$, $\text{skew}_{min}$, $\text{dunn}_{h1.5}$, $\text{silh}_{k1.5}$  \\ \hline ivy      & 0.432         & 0.636         & 0.568         & $\text{mfa}_{max}$                                                         & $\text{sd}_{min/max}$, $\text{silh}_{k1.5}$                            \\ \hline jedit    & 0.405         & 0.548         & 0.500         & $\text{ce}_{min}$, $\text{npm}_{sd}$                                           & $\text{dunn}_{k1.5}$                                            \\ \hline log4j    & 0.432         & 0.568         & 0.659         & $\text{rfc}_{med}$, $\text{ic}_{med}$                                    & $\text{skew}_{mean}$, $\text{kurt}_{mean}$, $\text{silh}_{k1.5}$                \\ \hline lucene   & 0.409         & 0.545         & 0.591         & $\text{avg\_cc}_{med}$                                                  & $\text{sd}_{min/max}$, $\text{skew}_{min}$, $\text{conn}_{k2}$                    \\ \hline pbeans   & 0.444         & 0.533         & 0.622         & $\text{dit}_{min}$                                                         & $\text{skew}_{mean}$, $\text{silh}_{h2}$, $\text{silh}_{k1.5}$          \\ \hline poi      & 0.442         & 0.698         & 0.535         & $\text{max\_cc}_{max}$, $\text{dit}_{min}$, $\text{avg\_cc}_{sd}$                  & $\text{mean}_{min/max}$                                             \\ \hline synapse  & 0.477         & 0.682         & 0.636         & $\text{avg\_cc}_{sd}$, $\text{moa}_{med}$, $\text{max\_cc}_{med}$            & $\text{sd}_{min/max}$, $\text{corr}_{mean}$, $\text{skew}_{mean}$, $\text{silh}_{k1.5}$    \\ \hline tomcat   & 0.457         & 0.674         & 0.587         & $\text{npm}_{max}$, $\text{loc}_{min}$, $\text{avg\_cc}_{sd}$                      & $\text{silh}_{k1.5}$                                        \\ \hline velocity & 0.477         & 0.614         & 0.523         & $\text{cbm}_{max}$, $\text{moa}_{sd}$, $\text{ca}_{med}$                        & $\text{corr}_{max}$                                                \\ \hline xalan    & 0.395         & 0.674         & 0.488         & $\text{amc}_{max}$, $\text{lcom}_{mean}$, $\text{rfc}_{sd}$, $\text{avg\_cc}_{med}$ & $\text{kurt}_{mean}$                                               \\ \hline xerces   & 0.465         & 0.698         & 0.628         & $\text{loc}_{min}$, $\text{cam}_{min}$, $\text{ca}_{mean}$, $\text{avg\_cc}_{med}$  & $\text{sd}_{min/max}$, $\text{kurt}_{mean}$, $\text{silh}_{k1.5}$                \\ \hline Mean     & 0.451 & 0.629 & 0.602 & -                                                                      & -                                                        \\ \hline \end{tabular}
\end{table}

\begin{figure}[!t]
\centering
\includegraphics[scale=0.8]{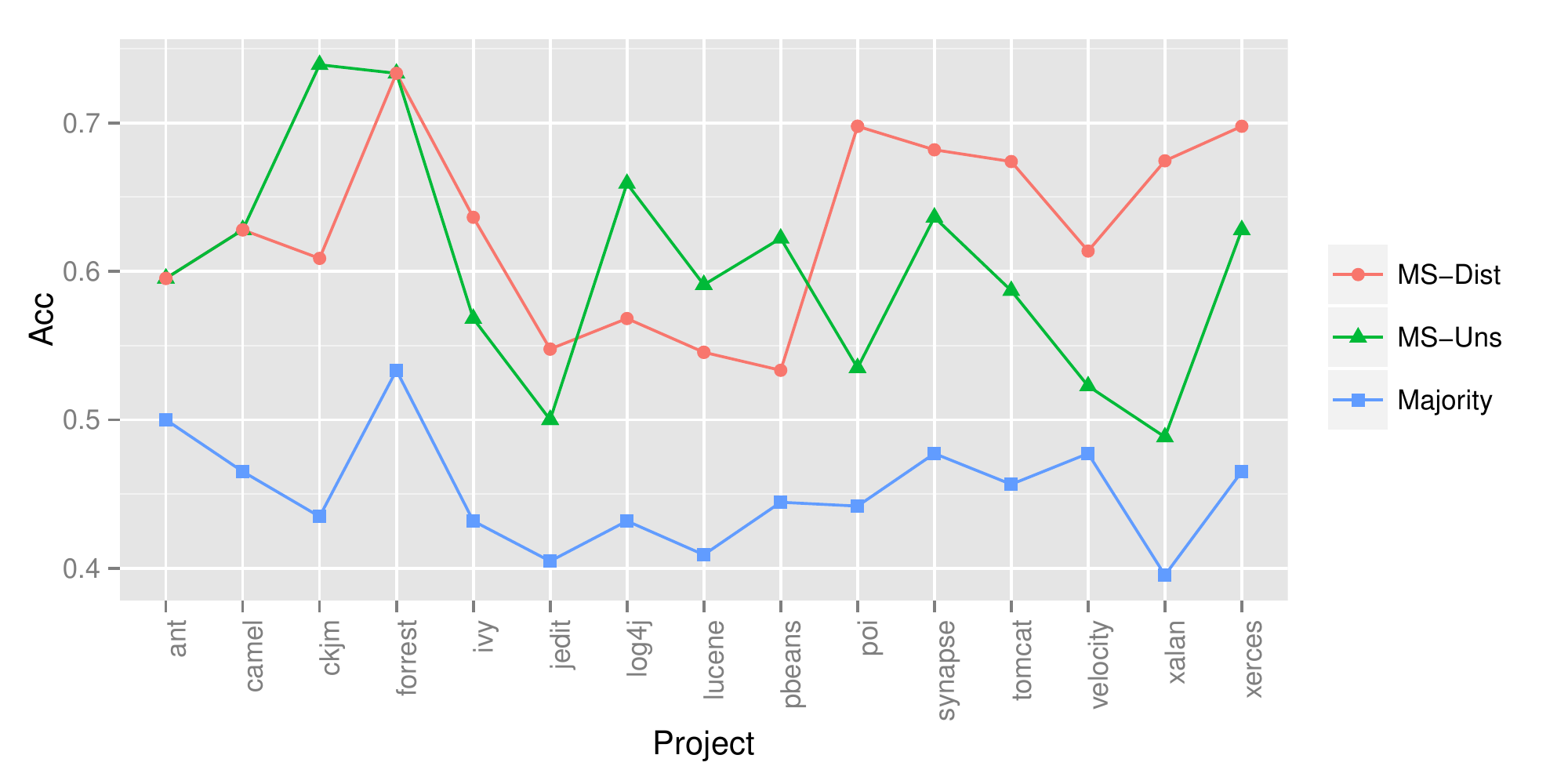}
\caption{\label{fig:Highest-subset-accuracy}The resulting best accuracies
generated for each meta-model in the attribute selection step. The
majority class is used as a comparative reference for the meta-attributes
sets MS-Dist and MS-Uns.}
\end{figure}

Initially, we can observe that both meta-attributes sets \emph{MS-Uns}
and \emph{MS-Dist} produced an accuracy superior to the majority class
for all projects meta-data. \emph{MS-Dist} produced the best mean
value although it cannot be differentiated from \emph{MS-Uns} with
statistical significance. These performances are better visualized
in Figure \ref{fig:Highest-subset-accuracy}.

Although the obtained performances characterize some level of meta-learning
capacity, they can also represent an overfitting on data; i.e., when
a subset achieves a high accuracy estimate but poor predictive power
for new examples \cite{Kohavi1995}. For small samples and a high
dimensionality of attributes (such as the evaluated context), it is
likely that one of the many attribute subsets lead to a hypothesis
with high predictive accuracy \cite{Kohavi1995}. 

The existence of overfitting can be verified by testing the meta-models
for new examples. For this, we use each meta-model to recommend a
suitable CPDP method for the respective project versions previously
separated for test (see Section \ref{sub:Experimental-Setup}). The
meta-learners \emph{MS-Dist} and \emph{MS-Uns} recommended correctly
a suitable CPDP method for 25 (53\%) and 16 (34\%) of the 47 tested
project versions, respectively. The majority class, represented by
the label \emph{2013He\_rf}, achieved the best AUC for 20 (43\%) of
all project versions. These results indicate an appropriate learning
of the meta-learning solution based on the meta-attributes set \emph{MS-Dist}
since it produced an accuracy superior to the majority class, defined
as the baseline.

\medskip{}

\begin{itemize}
\item \emph{RQ3.2: How does the meta-learner perform across datasets? (Base-level)}
\end{itemize}
\medskip{}

The general performance of the meta-learning solution is discussed
below. We compare the resulting recommendations provided in the meta-CPLOO
procedure in relation to the four evaluated CPDP methods applied individually.
We also compare the meta-learning performances with the random baseline.
Table \ref{tab:The-general-performance} presents the mean ranking
and mean AUC values for each method as well as the frequency in which
each method performed the best and worst AUC value for a project among
the four possible labels.

\begin{table}[!t]
\centering
\caption{\label{tab:The-general-performance}The General Performance of the
Proposed Meta-Learning Solution for Both Meta-Attributes Sets. These
Performances are Compared with the Four Base CPDP Methods and the
Random Baseline.}

\begin{tabular}{lcccc}\hline \textbf{Method}         & \textbf{Mean Rank}        & \textbf{Mean AUC}          & \textbf{Freq. Best} & \textbf{Freq. Worst} \\ \hline \textbf{Meta\_MS-Dist} & \textbf{3.40 ($\pm1.86$)} & \textbf{0.774 ($\pm0.10$)} & \textbf{25 (53\%)}  & \textbf{12 (26\%)}   \\ \hline 2012Ma\_nb              & 3.61 ($\pm1.88$)          & 0.771 ($\pm0.10$)          & 19 (40\%)           & 12 (26\%)            \\ \hline 2013He\_rf              & 3.76 ($\pm1.83$)          & 0.766 ($\pm0.09$)          & 20 (43\%)           & 11 (23\%)            \\ \hline \textbf{Meta\_MS-Uns}  & \textbf{4.03 ($\pm1.81$)} & \textbf{0.770 ($\pm0.09$)} & \textbf{16 (34\%)}  & \textbf{17 (36\%)}   \\ \hline \textbf{Random}         & \textbf{4.32 ($\pm0.92$)} & \textbf{0.766 ($\pm0.09$)} & \textbf{4 (9\%)}    & \textbf{3 (6\%)}     \\ \hline 2013He\_svm             & 4.37 ($\pm1.95$)          & 0.765 ($\pm0.09$)          & 13 (28\%)           & 18 (38\%)            \\ \hline 2009Turhan\_nb          & 4.51 ($\pm2.13$)          & 0.761 ($\pm0.09$)          & 12 (26\%)           & 21 (45\%)            \\ \hline \end{tabular} 
\end{table}

The meta-learner \emph{MS-Dist} presented the highest mean AUC of
all evaluated methods. This solution also presented the highest frequency
of best AUC. It provided the best AUC performance for 53\% of all
project versions against 40\% of the second best-ranked method (\emph{2012Ma\_nb}).
Together with \emph{2012Ma\_nb}, the solution \emph{MS-Dist} provided
the worst AUC performance for only 26\% of all project versions. It
is worth to note that even the worst AUC performance represents one
of the four best-ranked solution across datasets, as discussed in
Section \ref{sub:Meta-target}.

\begin{figure}[!t]
\centering
\includegraphics[scale=0.45]{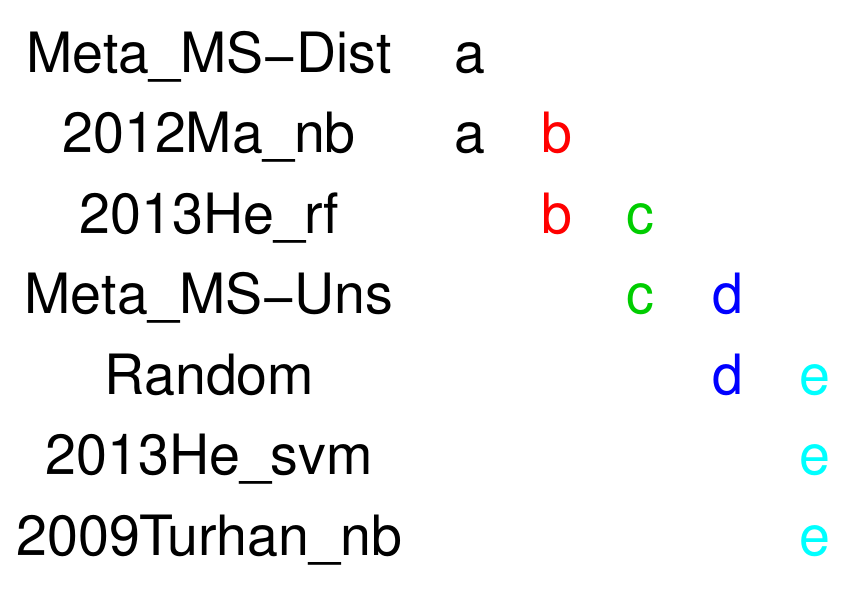}
\caption{\label{fig:meta_statistical}Pairwise Post-hoc Fisher’s LSD test.
The meta-learner MS-Dist presented the best ranking position although
it is not significantly different from the second best-ranked base
method \emph{2012Ma\_nb}. }
\end{figure}

We applied the Friedman test and obtained the \emph{p-value}$=$0.044,
which refuses the null hypothesis of performance equivalence between
methods. The results of the Fisher’s LSD test are presented in Figure
\ref{fig:meta_statistical}. From these results, we can highlight:
1) the random baseline is not significantly different from \emph{2013He\_svm}
and \emph{2009Turhan\_nb}; 2) the meta-learner \emph{MS-Uns} cannot
be differentiated from the random baseline; 3) the three best-ranked
methods \emph{MS-Dist}, \emph{2012Ma\_nb}, and \emph{2013He\_rf} present
significant difference in relation to the random baseline; and 4)
the two best-ranked methods do not present significant difference
from each other although \emph{MS-Dist} was more frequently the best approach (53\% against 40\%).

\section{Threats to Validity\label{sec:Threats-to-Validity}}

Some factors can threat the validity of the experiments conducted
in this study. The first issues are related to the collection of data
used in the experiments. Any lack of quality in these data may jeopardize
the entire study. For example, it is known that the number of defects
found for each software part is an approximate estimate and does not
represent the actual number of defects \cite{Madeyski2014}. However,
a precise information in this context is difficult if not impossible
to acquire in a real project \cite{Yang2015}. \citet{Kitchenham2007}
argue that software companies frequently do not keep suitable historical
information about software quality. When available, those information
are commonly private or restricted only for internal use \cite{Zimmermann2008}.
Also, the collection of data is composed only by Java open source
projects, which restrict the external validity of the results. 

Despite these factors, some positive points justify the use of this
collection of data: 1) the datasets are open for reuse; 2) the independent
variables can be automatically extracted; 3) the collection is composed
of several project versions, allowing us to conduct the experiments
in the CPDP context; 4) the data acquisition is based on a systematic
and frequently used process \cite{Zimmermann2007,Zhang2014,Madeyski2014};
and 5) these data were extensively reused in the literature \cite{Jureczko2010,He2013,Prateek2013,Herbold2013,Madeyski2014,Zhang2016}.

The methods evaluated in this study (including transfer learning solutions
and classifiers) compose a representative sample of the state-of-the-art
in CPDP. However, this does not represents an exhaustive comparison
covering all the existing solutions. The extension of these experiments
with additional methods can lead to alternative conclusions. Also,
the internal parameters of each method follow either the default configuration
provided by the code libraries or the original recommendations of
the authors. The tuning of parameters can influence the performances
and, consequently, the conclusions over results. Although the use
of default configurations is a common approach in the experimental
software engineering literature, future work should investigate the
impact of parameter tuning on this analysis. 

The preprocessing step also influences directly on the performance
of predictive models \cite{Keung2013}. We set the same log transformation
resource for all methods in order to diminish the impact of this step
on results. However, some internal preprocessing resources can still
interfere. For example, the Naive Bayes classifier operates in conjunction
with data discretization; which may benefit this classifier over other
\cite{Keung2013}. Also, \citet{Herbold2013} argues that the performance
of SVMs can be positively impacted by the weighting of imbalanced
training data, which is ignored in this study.

Another important issue, specific for this study, regards to the CPDP
methods considered as the labels for the meta-target. Those CPDP methods
presented the best ranking performances across datasets. Also, they
presented distinct bias in relation to each other, which can contribute
both for the diversity as well as the complementarity of the meta-learning
solution. However, other criteria can be considered in this case,
which may completely alter the obtained conclusions.

\section{Conclusions\label{sec:Conclusions}}

In this study we provided two main contributions. First, we conducted
an experimental comparison of 31 CPDP methods derived from six state-of-the-art
transfer learning solutions associated to the five most frequently
used classifiers in the defect prediction literature. Second, we investigated
the feasibility of meta-learning applied to CPDP. 

The experiments are based on 47 versions of 15 open source software
projects. Different from previous studies, we considered a context in which 
no previous information about the software is demanded.
In practice, this characteristic allows a broader applicability of
defect prediction models and the proposed meta-learning solution.
For example, for companies with insufficient data from previous projects
or for projects in its first release. This characteristic is possible
due to three main factors: 1) the training set is composed of known
external projects; 2) the software characterization can be extracted
directly from the source code; and 3) the meta-model is constructed
based on unsupervised meta-attributes.

From the first experiment, we identified the four best-ranked CPDP
methods across datasets in terms of AUC: \emph{2012Ma\_nb}, \emph{2013He\_rf},
\emph{2009Turhan\_nb}, and \emph{2013He\_svm}. These four methods
did not present significant difference of performance in relation
to each other. These four methods, however, presented the best performance
for distinct groups of datasets. In other words, the most suitable
CPDP method for a project varies according to the project being predicted.
These results accredited the investigation of the meta-learning solution
proposed in this study. 

We evaluated two distinct unsupervised meta-attributes sets for a
multi-label meta-learning task. The performance analysis was conducted
in two levels. In the meta-level, the results indicate a proper predictive
power of the proposed solution. The meta-learner based on the meta-attributes
set \emph{MS-Dist} presented an accuracy 10 percentage points superior
to the majority class. In the base-level, we compared the proposed
solution in relation to the four best-ranked CPDP methods across datasets
in terms of AUC. The meta-learner \emph{MS-Dist} presented the higher
mean AUC although it did not present significant difference of performance
in relation to the base method \emph{2012Ma\_nb}. 

Assuming the proper generalization of these results, three factors
contribute to the practical use of the proposed solution: 1) there
is learning in the meta-level. The meta-learner provided the best
solution for a larger amount of datasets than each of the four evaluated
methods applied individually; 2) in the worst case, the meta-learner
will still recommend one of the four CPDP methods with best ranking
across datasets; and 3) considering its application domain, the proposed
solution is not expensive. The hard computational cost is spent in
the meta-model construction. For the recommendation task, only the
meta-characterization and prediction costs are demanded.

However, further studies are still necessary to guarantee the generalization
of the presented results. In addition, alternative solutions can be
investigated aiming to improve the meta-learning performance. We point
out three factors for future investigation. First, the meta-data can
be expanded. A larger amount of examples can contribute to improve
the predictive power of the meta-learner. Second, other meta-attributes
sets can be explored. For example, the relation (or difference) between
testing and training data can be considered in conjunction with the
particularities of each dataset. Third, other attribute selection
methods can lead to subsets with higher predictive power.


%

\section*{Acknowledgments}
Research developed with the computational resources from the Center
for Mathematical Sciences Applied to Industry (CeMEAI) financed by
FAPESP. The authors also acknowledge the support granted by FAPESP
(grant 2013/01084-3).




\bibliographystyle{IEEEtranN}
\bibliography{references}
%

\begin{IEEEbiography}[{\includegraphics[width=1in,height=1.25in,clip,keepaspectratio]{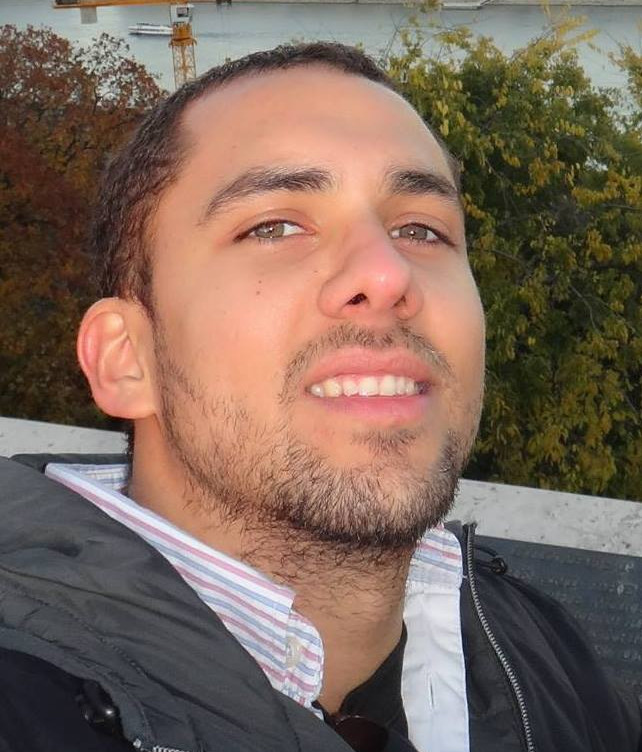}}]{Faimison Porto}
Faimison Porto received the BS degree in Computer Science from the Federal University of Goias (UFG), Brazil, in 2011, and the MS and PhD degrees in Computer Science from the University of Sao Paulo (USP), Brazil, in 2013 and 2017, respectively. From July 2016 to December 2016 he has worked in a research internships abroad program in collaboration with the University of Oulu (Finland) and the Blekinge Institute of Technology (Sweden). Currently, he is a lecturer at the Department of Computing and Electronics, Federal University of Espirito Santo (UFES), Brazil. In his academic career, he has worked with formal software testing and software defect prediction.
\end{IEEEbiography}

\begin{IEEEbiography}[{\includegraphics[width=1in,height=1.25in,clip,keepaspectratio]{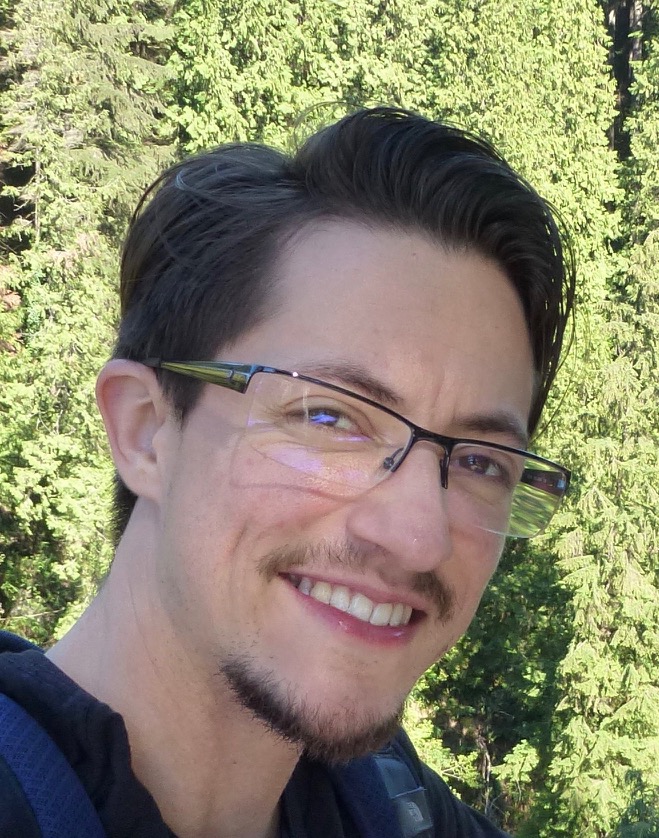}}]{Leandro Minku}
Dr. Leandro L. Minku is a Lecturer in Computer Science at the Department of Informatics, University of Leicester (UK). Prior to that, he was a research fellow at the University of Birmingham (UK) for five years. He received the PhD degree in Computer Science from the University of Birmingham (UK) in 2010. During his PhD, he was the recipient of the Overseas Research Students Award (ORSAS) from the British government and was invited to a 6-month internship at Google. Dr. Minku's main research interests are machine learning for software engineering, search-based software engineering, machine learning for non-stationary environments / data stream mining, and ensembles of learning machines. His work has been published in internationally renowned journals such as IEEE Transactions on Software Engineering, ACM Transactions on Software Engineering and Methodology, IEEE Transactions on Knowledge and Data Engineering, and IEEE Transactions on Neural Networks and Learning Systems. Among other roles, Dr. Minku is a steering committee member for the International Conference on Predictive Models and Data Analytics in Software Engineering (PROMISE), an associate editor for the Journal of Systems and Software, a conference correspondent for IEEE Software, and was the co-chair for the challenge track of the Symposium on Search-Based Software Engineering (SSBSE'2015-2016).
\end{IEEEbiography}

\begin{IEEEbiography}[{\includegraphics[width=1in,height=1.25in,clip,keepaspectratio]{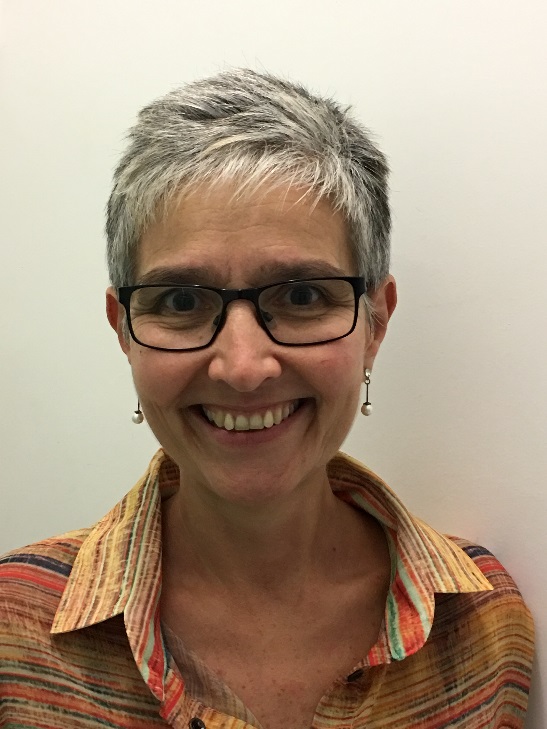}}]{Emilia Mendes}
Emilia Mendes is Full Professor in Computer Science at the Blekinge Institute of Technology (Sweden), and also a Tekes-funded Finnish Distinguished Professor at the University of Oulu (Finland). Her areas of research interest are mainly within the context of empirical software engineering, value-based software engineering, and the use of machine learning techniques to contexts such as healthcare, and sustainability. She has published widely and over 200 refereed papers, plus two books as solo author – both in the area of cost estimation. She is on the editorial board of several journals, which include TSE and the SQJ.
\end{IEEEbiography}

\begin{IEEEbiography}[{\includegraphics[width=1in,height=1.25in,clip,keepaspectratio]{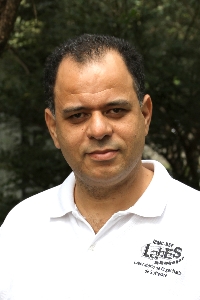}}]{Adenilso Simao}
Adenilso Simao received the BS degree in computer science from the State University of Maringa (UEM), Brazil, in 1998, and the MS and PhD degrees in computer science from the University of Sao Paulo (USP), Brazil, in 2000 and 2004, respectively. Since 2004, he has been a professor of computer science at the Computer System Department of USP. From August 2008 to July 2010, he has been on a sabbatical leave at Centre de Recherche Informatique de Montreal (CRIM), Canada. He has received best paper awards in several important conferences. He has also received distinguishing teacher awards in many occasions. His research interests include software testing and formal methods.
\end{IEEEbiography}
%
%
%



\pagebreak 

\newpage
\appendices
\section{\label{tab:code_metrics}Code Metrics}

\begin{itemize}

\item Average Method Complexity (AMC): This metric measures the average method size for each class. Size of a method is equal to the number of Java binary codes in the method.                                                                                                                                                                                                                                                                                                                                                                                                                                                                                 \item Cohesion Among Class Methods (CAM)      : This metric computes the relatedness among methods of a class based upon the parameter list of the methods. The metric is computed using the summation of number of different types of method parameters in every method divided by a multiplication of number of different method parameter types in whole class and number of methods.                                                                                                                                                                                                                                                                                 \item Afferent couplings (Ca): The Ca metric represents the number of classes that depend upon the measured class.                                                                                                                                                                                                                                                                                                                                                                                                                                                                                                                                      \item Efferent couplings (Ce): The Ce metric represents the number of classes that the measured class is depended upon.                                                                                                                                                                                                                                                                                                                                                                                                                                                                                                                                 \item Coupling Between Methods (CBM): The metric measures the total number of new/redefined methods to which all the inherited methods are coupled. There is a coupling when at least one of the conditions given in the IC metric is held.                                                                                                                                                                                                                                                                                                                                                                                                                    \item Coupling between object classes (CBO): The CBO metric represents the number of classes coupled to a given class (efferent couplings and afferent couplings).                                                                                                                                                                                                                                                                                                                                                                                                                                                                                                    \item 
Cyclomatic Complexity (CC):  CC is equal to number of different paths in a method (function) plus one. The McCabe cyclomatic complexity is defined as: CC=E-N+P; where E is the number of edges of the graph, N is the number of nodes of the graph, and P is the number of connected components. CC is the only method size metric. The constructed models make the class size predictions. Therefore, the metric had to be converted to a class size metric. Two metrics has been derived: 
\begin{itemize}
\item MAX\_CC - the greatest value of CC among methods of the investigated class;
\item AVG\_CC - the arithmetic mean of the CC value in the investigated class.
\end{itemize}

Data Access Metric (DAM): This metric is the ratio of the number of private (protected) attributes to the total number of attributes declared in the class.                                                                                                                                                                                                                                                                                                                                                                                                                                                                                        \item Depth of Inheritance Tree (DIT): The DIT metric provides for each class a measure of the inheritance levels from the object hierarchy top.                                                                                                                                                                                                                                                                                                                                                                                                                                                                                                                \item Inheritance Coupling (IC): This metric provides the number of parent classes to which a given class is coupled. A class is coupled to its parent class if one of its inherited methods functionally dependent on the new or redefined methods in the class. A class is coupled to its parent class if one of the following conditions is satisfied:
\begin{itemize}
\item One of its inherited methods uses an attribute that is defined in a new/redefined method;
\item One of its inherited methods calls a redefined method; 
\item One of its inherited methods is called by a redefined method and uses a parameter that is defined in the redefined method.               
\end{itemize}
\item Lack of cohesion in methods (LCOM): The LCOM metric counts the sets of methods in a class that are not related through the sharing of some of the class fields.                                                                                                                                                                                                                                                                                                                                                                                                                                                                                              \item Lack of cohesion in methods (LCOM3): 

$LCOM3=\frac{\left(\frac{1}{a}\sum_{j=1}^{a}\mu(A_{j})\right)-m}{1-m}$

\begin{itemize}
\item m - number of methods in a class;
\item a - number of attributes in a class;
\item $\mu(A)$ - number of methods that access the attribute A.
\end{itemize}  
 \item Lines of Code (LOC): The LOC metric calculates the number of lines of code in the Java binary code of the class under investigation.                                                                                                                                                                                                                                                                                                                                                                                                                                                                                                          \item Measure of Functional Abstraction (MFA): This metric is the ratio of the number of methods inherited by a class to the total number of methods accessible by the member methods of the class.                                                                                                                                                                                                                                                                                                                                                                                                                                                                     \item Measure of Aggregation (MOA): This metric measures the extent of the part-whole relationship, realized by using attributes. The metric is a count of the number of class fields whose types are user defined classes.                                                                                                                                                                                                                                                                                                                                                                                                                                  \item Number of Children (NOC): The NOC metric simply measures the number of immediate descendants of the class.                                                                                                                                                                                                                                                                                                                                                                                                                                                                                                                                         \item Number of Public Methods (NPM): The NPM metric counts all the methods in a class that are declared as public.                                                                                                                                                                                                                                                                                                                                                                                                                                                                                                                                            \item Response for a Class (RFC): The RFC metric measures the number of different methods that can be executed when an object of that class receives a message.                                                                                                                                                                                                                                                                                                                                                                                                                                                                                            \item Weighted methods per class (WMC): The value of the WMC is equal to the number of methods in the class (assuming unity weights for all methods).                                                                                                                              \end{itemize}                                                                                                                                                                                                                                                                                                                                                                                 
For a detailed explanation see \cite{Jureczko2010}.

\section{\label{tab:five-classifiers}Classifiers}

\begin{itemize}

\item Random Forest (RF): RF is based on a collection of decision trees where each tree
is grown from a bootstrap sample (randomly sampling the data with
replacement). The process of finding the best split for each node
is based on a subset of attributes randomly chosen. This characteristic
produces a collection of trees with different biases. The final prediction
for a new example is given by the majority voting of all trees. This
algorithm is robust to redundant and irrelevant attributes although
it can produce overfitting models.
\item Support Vector Machines (SVM): SVM aims to find the optimal hyperplane that maximally separates
samples in two different classes. This classifier is also robust to
redundant and irrelevant attributes since the number of attributes
does not affect the complexity of an SVM model. In this study, we
evaluate a SVM variation called Sequential Minimal Optimization (SMO).
This technique optimizes the SVM training by dividing the large Quadratic
Programming (QP) problem in a series of possible QP problems \cite{Platt1999}.
\item Multilayer Perceptron (MLP): MLP is a neural network model based on the back-propagation
algorithm. The MLP consists of three or more layers (an input and
an output layer with one or more hidden layers) of nodes in a directed
graph. Each layer is fully connected to the next one. The nodes (except
for the input nodes) are neurons (or processing elements) with a nonlinear
activation function. The weights of each node in the network are iteratively
updated in an attempt to minimize a loss function calculated from
the output layers.
\item C4.5 (C45): C45 is a decision tree algorithm which extends the Quinlan's
earlier ID3 algorithm. A decision tree is a collection of decision
rules defined in each node. The tree is generated by associating to
each node the attribute that most effectively divides the set of training
data. The splitting criterion is the normalized information gain \cite{Maimon2010}.
The classification of examples is performed by following a path trough
the tree from the root to the leaf nodes where a class value is taken.
\item Naive Bayes (NB): NB is a simple statistical algorithm based on Bayes' Theorem.
In the defect prediction context, this theorem can be defined as follows:
\[
P(c_{k}|x)=\frac{P(c_{k})}{P(x)}\prod_{j=1}^{m}P(x_{j}|c_{k})
\]
where $c_{k}$ is an element of the set of class values (defective
or not-defective), $x=(x_{1},...,x_{m})$ is an attributes vector,
$P(c_{k})$ and $P(x)$ are respectively the prior probabilities of
$c_{k}$ and $x$ occur, and $P(x_{j}|c_{k})$ is the probability
of $x_{j}$ given $c_{k}$. These probabilities are combined based
on the training set. The theorem calculates the posterior probability
of $c_{k}$ given that $x$ is true. The term ``naïve'' is due to
its assumption that the attributes are independent. Although this
assumption is not always true, this algorithm has been reported as
an efficient classifier for defect prediction \cite{Lessmann2008}.

\end{itemize}

For a detailed explanation see \cite{Maimon2010,Sammut2017}.

\section{General AUC Performances}
\begin{table*}[!t]
\renewcommand{\arraystretch}{1.3}
\centering
\caption{\label{tab:General-performance-II}Part I - General AUC Performances
for All the Evaluated CPDP Methods (Base and Meta). The Methods are
Ordered by the Mean AUC Value. }

\resizebox{\textwidth}{!}{
\rotatebox{270}{
\begin{tabular}{lccccccccccccccccc} 
\textbf{Project}  & \rot{\textbf{Meta\_MS-Dist}} & \rot{\textbf{2012Ma\_nb}} & \rot{\textbf{Meta\_MS-Uns}} & \rot{\textbf{2013He\_rf}} & \rot{\textbf{Random}} & \rot{\textbf{2013He\_svm}} & \rot{\textbf{2009Turhan\_nb}} & \rot{\textbf{2013Herbold\_nb}} & \rot{\textbf{orig\_nb}} & \rot{\textbf{2009Cruz\_nb}} & \rot{\textbf{orig\_rf}} & \rot{\textbf{2013He\_nb}} & \rot{\textbf{2013He\_mlp}} & \rot{\textbf{2008Watanabe\_rf}} & \rot{\textbf{2008Watanabe\_nb}} & \rot{\textbf{2013Herbold\_rf}} & \rot{\textbf{2009Turhan\_rf}} \\ \hline ant-1.3           & 0.879                  & 0.879               & 0.860                 & 0.870               & 0.870           & 0.860                & 0.872                   & 0.870                    & 0.867             & 0.867                 & 0.806             & 0.873               & 0.838                & 0.859                     & 0.867                     & 0.848                    & 0.826                   \\ \hline ant-1.4           & 0.679                  & 0.679               & 0.663                 & 0.663               & 0.664           & 0.656                & 0.656                   & 0.658                    & 0.662             & 0.660                 & 0.677             & 0.622               & 0.640                & 0.682                     & 0.658                     & 0.693                    & 0.686                   \\ \hline ant-1.5           & 0.753                  & 0.836               & 0.836                 & 0.780               & 0.801           & 0.753                & 0.821                   & 0.813                    & 0.814             & 0.818                 & 0.767             & 0.780               & 0.752                & 0.814                     & 0.816                     & 0.796                    & 0.739                   \\ \hline ant-1.6           & 0.829                  & 0.858               & 0.829                 & 0.854               & 0.848           & 0.829                & 0.849                   & 0.849                    & 0.846             & 0.842                 & 0.846             & 0.851               & 0.801                & 0.845                     & 0.845                     & 0.851                    & 0.813                   \\ \hline ant-1.7           & 0.824                  & 0.851               & 0.851                 & 0.832               & 0.835           & 0.824                & 0.842                   & 0.840                    & 0.839             & 0.839                 & 0.791             & 0.819               & 0.778                & 0.813                     & 0.838                     & 0.802                    & 0.778                   \\ \hline camel-1.0         & 0.830                  & 0.783               & 0.783                 & 0.782               & 0.804           & 0.830                & 0.837                   & 0.827                    & 0.821             & 0.825                 & 0.758             & 0.805               & 0.763                & 0.741                     & 0.802                     & 0.744                    & 0.732                   \\ \hline camel-1.2         & 0.622                  & 0.629               & 0.629                 & 0.641               & 0.632           & 0.622                & 0.634                   & 0.624                    & 0.626             & 0.625                 & 0.648             & 0.629               & 0.639                & 0.647                     & 0.625                     & 0.643                    & 0.647                   \\ \hline camel-1.4         & 0.692                  & 0.729               & 0.729                 & 0.714               & 0.720           & 0.692                & 0.741                   & 0.731                    & 0.735             & 0.734                 & 0.734             & 0.717               & 0.681                & 0.735                     & 0.739                     & 0.704                    & 0.736                   \\ \hline camel-1.6         & 0.673                  & 0.650               & 0.650                 & 0.678               & 0.671           & 0.673                & 0.666                   & 0.666                    & 0.664             & 0.666                 & 0.643             & 0.679               & 0.663                & 0.673                     & 0.657                     & 0.660                    & 0.640                   \\ \hline ckjm-1.8          & 0.960                  & 0.920               & 0.920                 & 0.720               & 0.851           & 0.960                & 0.720                   & 0.880                    & 0.880             & 0.880                 & 0.680             & 0.840               & 0.960                & 0.960                     & 0.920                     & 0.640                    & 0.560                   \\ \hline forrest-0.7       & 0.808                  & 0.762               & 0.762                 & 0.792               & 0.790           & 0.792                & 0.808                   & 0.738                    & 0.715             & 0.708                 & 0.915             & 0.785               & 0.762                & 0.831                     & 0.685                     & 0.877                    & 0.923                   \\ \hline forrest-0.8       & 0.906                  & 0.859               & 0.859                 & 0.812               & 0.854           & 0.906                & 0.828                   & 0.719                    & 0.734             & 0.703                 & 0.953             & 0.781               & 0.938                & 0.766                     & 0.656                     & 0.703                    & 0.953                   \\ \hline ivy-1.1           & 0.814                  & 0.812               & 0.814                 & 0.814               & 0.791           & 0.775                & 0.759                   & 0.759                    & 0.755             & 0.747                 & 0.714             & 0.757               & 0.720                & 0.790                     & 0.747                     & 0.728                    & 0.718                   \\ \hline ivy-1.4           & 0.805                  & 0.779               & 0.768                 & 0.805               & 0.783           & 0.783                & 0.768                   & 0.793                    & 0.780             & 0.782                 & 0.772             & 0.781               & 0.763                & 0.790                     & 0.778                     & 0.746                    & 0.798                   \\ \hline ivy-2.0           & 0.846                  & 0.851               & 0.846                 & 0.846               & 0.843           & 0.839                & 0.835                   & 0.831                    & 0.831             & 0.830                 & 0.815             & 0.825               & 0.797                & 0.812                     & 0.818                     & 0.796                    & 0.784                   \\ \hline jedit-3.2.1       & 0.878                  & 0.887               & 0.879                 & 0.878               & 0.880           & 0.878                & 0.879                   & 0.880                    & 0.876             & 0.876                 & 0.881             & 0.880               & 0.850                & 0.872                     & 0.878                     & 0.873                    & 0.887                   \\ \hline jedit-4.0         & 0.874                  & 0.860               & 0.860                 & 0.874               & 0.865           & 0.862                & 0.866                   & 0.864                    & 0.858             & 0.859                 & 0.856             & 0.854               & 0.855                & 0.833                     & 0.860                     & 0.867                    & 0.852                   \\ \hline jedit-4.1         & 0.908                  & 0.900               & 0.900                 & 0.908               & 0.899           & 0.906                & 0.887                   & 0.891                    & 0.889             & 0.895                 & 0.876             & 0.886               & 0.896                & 0.893                     & 0.888                     & 0.883                    & 0.882                   \\ \hline jedit-4.2         & 0.924                  & 0.924               & 0.916                 & 0.924               & 0.922           & 0.916                & 0.922                   & 0.923                    & 0.922             & 0.922                 & 0.907             & 0.906               & 0.911                & 0.905                     & 0.918                     & 0.916                    & 0.909                   \\ \hline jedit-4.3         & 0.877                  & 0.877               & 0.881                 & 0.881               & 0.872           & 0.900                & 0.849                   & 0.847                    & 0.850             & 0.853                 & 0.854             & 0.828               & 0.849                & 0.850                     & 0.838                     & 0.820                    & 0.847                   \\ \hline log4j-1.0         & 0.888                  & 0.883               & 0.883                 & 0.872               & 0.830           & 0.743                & 0.888                   & 0.887                    & 0.877             & 0.881                 & 0.891             & 0.856               & 0.800                & 0.887                     & 0.880                     & 0.903                    & 0.813                   \\ \hline log4j-1.1         & 0.810                  & 0.858               & 0.851                 & 0.810               & 0.830           & 0.797                & 0.851                   & 0.838                    & 0.837             & 0.840                 & 0.806             & 0.816               & 0.784                & 0.786                     & 0.826                     & 0.827                    & 0.789                   \\ \hline log4j-1.2         & 0.881                  & 0.881               & 0.790                 & 0.826               & 0.837           & 0.790                & 0.820                   & 0.799                    & 0.814             & 0.845                 & 0.879             & 0.830               & 0.682                & 0.864                     & 0.814                     & 0.878                    & 0.867                   \\ \hline lucene-2.0        & 0.792                  & 0.808               & 0.808                 & 0.792               & 0.789           & 0.775                & 0.778                   & 0.780                    & 0.777             & 0.773                 & 0.789             & 0.778               & 0.734                & 0.790                     & 0.775                     & 0.748                    & 0.792                   \\ \hline lucene-2.2        & 0.749                  & 0.735               & 0.728                 & 0.749               & 0.741           & 0.753                & 0.728                   & 0.733                    & 0.726             & 0.740                 & 0.708             & 0.736               & 0.736                & 0.724                     & 0.733                     & 0.712                    & 0.721                   \\ \hline lucene-2.4        & 0.804                  & 0.783               & 0.783                 & 0.804               & 0.793           & 0.801                & 0.788                   & 0.797                    & 0.793             & 0.800                 & 0.754             & 0.789               & 0.799                & 0.764                     & 0.795                     & 0.760                    & 0.720                   \\ \hline pbeans-1.0        & 0.612                  & 0.612               & 0.642                 & 0.642               & 0.640           & 0.650                & 0.675                   & 0.679                    & 0.658             & 0.654                 & 0.746             & 0.608               & 0.842                & 0.700                     & 0.683                     & 0.754                    & 0.740                   \\ \hline pbeans-2.0        & 0.758                  & 0.758               & 0.811                 & 0.758               & 0.780           & 0.781                & 0.811                   & 0.791                    & 0.792             & 0.796                 & 0.757             & 0.781               & 0.591                & 0.778                     & 0.794                     & 0.712                    & 0.798                   \\ \hline poi-1.5           & 0.758                  & 0.758               & 0.746                 & 0.709               & 0.737           & 0.746                & 0.747                   & 0.760                    & 0.765             & 0.730                 & 0.706             & 0.731               & 0.653                & 0.721                     & 0.739                     & 0.650                    & 0.677                   \\ \hline poi-2.0RC1        & 0.701                  & 0.635               & 0.635                 & 0.701               & 0.670           & 0.704                & 0.680                   & 0.658                    & 0.643             & 0.691                 & 0.680             & 0.678               & 0.732                & 0.744                     & 0.668                     & 0.683                    & 0.674                   \\ \hline poi-2.5.1         & 0.786                  & 0.786               & 0.766                 & 0.742               & 0.766           & 0.774                & 0.766                   & 0.764                    & 0.773             & 0.725                 & 0.739             & 0.735               & 0.721                & 0.651                     & 0.733                     & 0.726                    & 0.684                   \\ \hline poi-3.0           & 0.849                  & 0.845               & 0.849                 & 0.849               & 0.847           & 0.849                & 0.835                   & 0.829                    & 0.829             & 0.808                 & 0.759             & 0.829               & 0.840                & 0.711                     & 0.822                     & 0.752                    & 0.780                   \\ \hline synapse-1.0       & 0.777                  & 0.785               & 0.777                 & 0.777               & 0.762           & 0.752                & 0.746                   & 0.773                    & 0.779             & 0.770                 & 0.815             & 0.751               & 0.762                & 0.688                     & 0.696                     & 0.736                    & 0.786                   \\ \hline synapse-1.1       & 0.676                  & 0.682               & 0.681                 & 0.676               & 0.677           & 0.681                & 0.659                   & 0.680                    & 0.668             & 0.666                 & 0.640             & 0.679               & 0.690                & 0.576                     & 0.617                     & 0.668                    & 0.539                   \\ \hline synapse-1.2       & 0.747                  & 0.747               & 0.717                 & 0.717               & 0.735           & 0.706                & 0.769                   & 0.748                    & 0.756             & 0.751                 & 0.691             & 0.755               & 0.708                & 0.677                     & 0.761                     & 0.723                    & 0.705                   \\ \hline tomcat-6.0.389418 & 0.818                  & 0.818               & 0.818                 & 0.810               & 0.805           & 0.784                & 0.812                   & 0.817                    & 0.810             & 0.813                 & 0.805             & 0.780               & 0.760                & 0.801                     & 0.805                     & 0.807                    & 0.778                   \\ \hline velocity-1.4      & 0.584                  & 0.584               & 0.623                 & 0.623               & 0.609           & 0.625                & 0.580                   & 0.566                    & 0.560             & 0.551                 & 0.552             & 0.579               & 0.683                & 0.608                     & 0.542                     & 0.590                    & 0.556                   \\ \hline velocity-1.5      & 0.706                  & 0.712               & 0.712                 & 0.701               & 0.691           & 0.665                & 0.706                   & 0.712                    & 0.711             & 0.708                 & 0.630             & 0.715               & 0.638                & 0.632                     & 0.695                     & 0.643                    & 0.618                   \\ \hline velocity-1.6.1    & 0.728                  & 0.728               & 0.728                 & 0.734               & 0.722           & 0.756                & 0.685                   & 0.701                    & 0.696             & 0.696                 & 0.673             & 0.705               & 0.744                & 0.643                     & 0.667                     & 0.673                    & 0.657                   \\ \hline xalan-2.4.0       & 0.804                  & 0.791               & 0.791                 & 0.810               & 0.798           & 0.804                & 0.788                   & 0.783                    & 0.784             & 0.791                 & 0.761             & 0.782               & 0.779                & 0.750                     & 0.788                     & 0.761                    & 0.751                   \\ \hline xalan-2.5.0       & 0.646                  & 0.646               & 0.684                 & 0.684               & 0.657           & 0.662                & 0.628                   & 0.634                    & 0.632             & 0.638                 & 0.677             & 0.629               & 0.676                & 0.671                     & 0.635                     & 0.660                    & 0.682                   \\ \hline xalan-2.6.0       & 0.762                  & 0.743               & 0.743                 & 0.762               & 0.743           & 0.740                & 0.729                   & 0.728                    & 0.725             & 0.721                 & 0.740             & 0.724               & 0.760                & 0.752                     & 0.741                     & 0.725                    & 0.736                   \\ \hline xalan-2.7.0       & 0.852                  & 0.762               & 0.852                 & 0.852               & 0.778           & 0.807                & 0.697                   & 0.712                    & 0.712             & 0.717                 & 0.840             & 0.757               & 0.868                & 0.864                     & 0.733                     & 0.830                    & 0.845                   \\ \hline xerces-1.2.0      & 0.483                  & 0.517               & 0.529                 & 0.529               & 0.512           & 0.506                & 0.483                   & 0.488                    & 0.486             & 0.485                 & 0.563             & 0.527               & 0.573                & 0.495                     & 0.470                     & 0.648                    & 0.554                   \\ \hline xerces-1.3.0      & 0.725                  & 0.696               & 0.644                 & 0.644               & 0.683           & 0.653                & 0.725                   & 0.698                    & 0.705             & 0.706                 & 0.704             & 0.661               & 0.657                & 0.655                     & 0.728                     & 0.681                    & 0.711                   \\ \hline xerces-1.4.4      & 0.728                  & 0.728               & 0.695                 & 0.695               & 0.740           & 0.756                & 0.759                   & 0.732                    & 0.747             & 0.757                 & 0.717             & 0.717               & 0.741                & 0.741                     & 0.764                     & 0.656                    & 0.729                   \\ \hline xerces-init       & 0.589                  & 0.645               & 0.680                 & 0.680               & 0.632           & 0.620                & 0.589                   & 0.618                    & 0.602             & 0.606                 & 0.631             & 0.683               & 0.607                & 0.604                     & 0.590                     & 0.666                    & 0.616                   \\ \hline Mean              & 0.774                  & 0.771               & 0.770                 & 0.766               & 0.766           & 0.765                & 0.761                   & 0.760                    & 0.758             & 0.757                 & 0.756             & 0.756               & 0.753                & 0.753                     & 0.752                     & 0.748                    & 0.745                   \\ \hline SD                & 0.101                  & 0.096               & 0.092                 & 0.087               & 0.088           & 0.093                & 0.093                   & 0.092                    & 0.093             & 0.094                 & 0.093             & 0.086               & 0.090                & 0.098                     & 0.098                     & 0.084                    & 0.100                   \\ \hline Median            & 0.792                  & 0.783               & 0.783                 & 0.780               & 0.783           & 0.775                & 0.768                   & 0.764                    & 0.773             & 0.757                 & 0.757             & 0.778               & 0.760                & 0.752                     & 0.761                     & 0.736                    & 0.739                   \\ \hline \end{tabular} }}
\end{table*}

\begin{table*}[!t]
\renewcommand{\arraystretch}{1.3}
\centering
\caption{\label{tab:General-performance-I}Part II - General AUC Performances
for All the Evaluated CPDP Methods (Base and Meta). The Methods are
Ordered by the Mean AUC Value. }

\resizebox{\textwidth}{!}{
\rotatebox{270}{
\begin{tabular}{lccccccccccccccccc} 

\textbf{Project}  & \rot{\textbf{orig\_mlp}} & \rot{\textbf{2009Cruz\_rf}} & \rot{\textbf{2013He\_c45}} & \rot{\textbf{2013Herbold\_mlp}} & \rot{\textbf{2009Cruz\_svm}} & \rot{\textbf{orig\_svm}} & \rot{\textbf{2008Watanabe\_mlp}} & \rot{\textbf{2008Watanabe\_svm}} & \rot{\textbf{2009Turhan\_svm}} & \rot{\textbf{2009Cruz\_mlp}} & \rot{\textbf{2009Turhan\_mlp}} & \rot{\textbf{2013Herbold\_svm}} & \rot{\textbf{orig\_c45}} & \rot{\textbf{2013Herbold\_c45}} & \rot{\textbf{2009Turhan\_c45}} & \rot{\textbf{2009Cruz\_c45}} & \rot{\textbf{2008Watanabe\_c45}} \\ \hline ant-1.3           & 0.807              & 0.862                 & 0.780                & 0.819                     & 0.778                  & 0.778              & 0.836                      & 0.782                      & 0.748                    & 0.787                  & 0.657                    & 0.818                     & 0.736              & 0.648                     & 0.574                    & 0.715                  & 0.697                      \\ \hline ant-1.4           & 0.706              & 0.672                 & 0.642                & 0.681                     & 0.665                  & 0.665              & 0.701                      & 0.657                      & 0.664                    & 0.642                  & 0.573                    & 0.682                     & 0.666              & 0.644                     & 0.617                    & 0.643                  & 0.531                      \\ \hline ant-1.5           & 0.715              & 0.792                 & 0.746                & 0.760                     & 0.715                  & 0.715              & 0.737                      & 0.734                      & 0.701                    & 0.723                  & 0.641                    & 0.785                     & 0.643              & 0.662                     & 0.545                    & 0.581                  & 0.707                      \\ \hline ant-1.6           & 0.802              & 0.830                 & 0.806                & 0.815                     & 0.795                  & 0.795              & 0.787                      & 0.798                      & 0.790                    & 0.788                  & 0.713                    & 0.835                     & 0.669              & 0.732                     & 0.658                    & 0.668                  & 0.718                      \\ \hline ant-1.7           & 0.789              & 0.808                 & 0.799                & 0.777                     & 0.803                  & 0.803              & 0.802                      & 0.806                      & 0.782                    & 0.760                  & 0.709                    & 0.812                     & 0.707              & 0.658                     & 0.574                    & 0.643                  & 0.697                      \\ \hline camel-1.0         & 0.783              & 0.752                 & 0.743                & 0.711                     & 0.820                  & 0.820              & 0.724                      & 0.782                      & 0.492                    & 0.684                  & 0.777                    & 0.389                     & 0.665              & 0.630                     & 0.750                    & 0.702                  & 0.463                      \\ \hline camel-1.2         & 0.655              & 0.625                 & 0.604                & 0.631                     & 0.652                  & 0.652              & 0.637                      & 0.648                      & 0.655                    & 0.620                  & 0.650                    & 0.459                     & 0.613              & 0.553                     & 0.587                    & 0.548                  & 0.585                      \\ \hline camel-1.4         & 0.726              & 0.730                 & 0.640                & 0.698                     & 0.754                  & 0.754              & 0.712                      & 0.760                      & 0.677                    & 0.698                  & 0.669                    & 0.504                     & 0.595              & 0.583                     & 0.641                    & 0.601                  & 0.553                      \\ \hline camel-1.6         & 0.659              & 0.658                 & 0.678                & 0.626                     & 0.627                  & 0.627              & 0.636                      & 0.621                      & 0.612                    & 0.615                  & 0.599                    & 0.546                     & 0.543              & 0.543                     & 0.512                    & 0.516                  & 0.475                      \\ \hline ckjm-1.8          & 0.840              & 0.880                 & 0.880                & 0.680                     & 0.840                  & 0.840              & 0.720                      & 0.800                      & 0.680                    & 0.520                  & 0.680                    & 0.720                     & 0.720              & 0.720                     & 0.500                    & 0.800                  & 0.500                      \\ \hline forrest-0.7       & 0.731              & 0.842                 & 0.623                & 0.862                     & 0.838                  & 0.838              & 0.700                      & 0.838                      & 0.808                    & 0.662                  & 0.854                    & 0.500                     & 0.885              & 0.938                     & 0.804                    & 0.785                  & 0.512                      \\ \hline forrest-0.8       & 0.797              & 0.594                 & 0.750                & 0.844                     & 0.656                  & 0.656              & 0.828                      & 0.516                      & 1.000                    & 0.594                  & 0.953                    & 0.641                     & 0.773              & 0.664                     & 0.961                    & 0.469                  & 0.539                      \\ \hline ivy-1.1           & 0.736              & 0.730                 & 0.706                & 0.756                     & 0.747                  & 0.747              & 0.790                      & 0.737                      & 0.795                    & 0.782                  & 0.698                    & 0.770                     & 0.647              & 0.620                     & 0.656                    & 0.640                  & 0.651                      \\ \hline ivy-1.4           & 0.738              & 0.729                 & 0.692                & 0.721                     & 0.722                  & 0.722              & 0.778                      & 0.729                      & 0.728                    & 0.769                  & 0.623                    & 0.694                     & 0.784              & 0.472                     & 0.553                    & 0.642                  & 0.720                      \\ \hline ivy-2.0           & 0.846              & 0.815                 & 0.780                & 0.793                     & 0.799                  & 0.799              & 0.792                      & 0.791                      & 0.825                    & 0.851                  & 0.802                    & 0.788                     & 0.718              & 0.647                     & 0.721                    & 0.729                  & 0.624                      \\ \hline jedit-3.2.1       & 0.826              & 0.856                 & 0.864                & 0.820                     & 0.881                  & 0.881              & 0.821                      & 0.885                      & 0.876                    & 0.819                  & 0.730                    & 0.827                     & 0.612              & 0.698                     & 0.715                    & 0.630                  & 0.690                      \\ \hline jedit-4.0         & 0.802              & 0.814                 & 0.853                & 0.818                     & 0.864                  & 0.864              & 0.793                      & 0.864                      & 0.875                    & 0.802                  & 0.827                    & 0.843                     & 0.687              & 0.784                     & 0.705                    & 0.704                  & 0.708                      \\ \hline jedit-4.1         & 0.852              & 0.878                 & 0.864                & 0.822                     & 0.858                  & 0.858              & 0.818                      & 0.864                      & 0.857                    & 0.835                  & 0.746                    & 0.845                     & 0.725              & 0.752                     & 0.719                    & 0.730                  & 0.725                      \\ \hline jedit-4.2         & 0.864              & 0.893                 & 0.921                & 0.900                     & 0.885                  & 0.885              & 0.827                      & 0.885                      & 0.906                    & 0.868                  & 0.864                    & 0.884                     & 0.857              & 0.780                     & 0.750                    & 0.803                  & 0.732                      \\ \hline jedit-4.3         & 0.814              & 0.827                 & 0.913                & 0.845                     & 0.897                  & 0.897              & 0.761                      & 0.890                      & 0.894                    & 0.688                  & 0.853                    & 0.893                     & 0.748              & 0.871                     & 0.823                    & 0.871                  & 0.710                      \\ \hline log4j-1.0         & 0.860              & 0.824                 & 0.843                & 0.799                     & 0.791                  & 0.791              & 0.805                      & 0.795                      & 0.713                    & 0.746                  & 0.663                    & 0.573                     & 0.762              & 0.724                     & 0.605                    & 0.636                  & 0.648                      \\ \hline log4j-1.1         & 0.825              & 0.773                 & 0.787                & 0.785                     & 0.811                  & 0.811              & 0.740                      & 0.801                      & 0.803                    & 0.734                  & 0.709                    & 0.540                     & 0.626              & 0.633                     & 0.600                    & 0.582                  & 0.654                      \\ \hline log4j-1.2         & 0.889              & 0.805                 & 0.763                & 0.860                     & 0.744                  & 0.744              & 0.852                      & 0.785                      & 0.652                    & 0.782                  & 0.784                    & 0.457                     & 0.609              & 0.756                     & 0.818                    & 0.717                  & 0.635                      \\ \hline lucene-2.0        & 0.724              & 0.774                 & 0.744                & 0.716                     & 0.636                  & 0.636              & 0.699                      & 0.639                      & 0.587                    & 0.701                  & 0.695                    & 0.642                     & 0.580              & 0.667                     & 0.675                    & 0.624                  & 0.616                      \\ \hline lucene-2.2        & 0.703              & 0.702                 & 0.689                & 0.710                     & 0.674                  & 0.674              & 0.698                      & 0.686                      & 0.540                    & 0.690                  & 0.668                    & 0.651                     & 0.534              & 0.629                     & 0.666                    & 0.587                  & 0.552                      \\ \hline lucene-2.4        & 0.729              & 0.751                 & 0.742                & 0.723                     & 0.676                  & 0.676              & 0.722                      & 0.687                      & 0.654                    & 0.696                  & 0.652                    & 0.618                     & 0.593              & 0.589                     & 0.695                    & 0.624                  & 0.574                      \\ \hline pbeans-1.0        & 0.550              & 0.477                 & 0.700                & 0.512                     & 0.667                  & 0.667              & 0.446                      & 0.675                      & 0.579                    & 0.554                  & 0.700                    & 0.596                     & 0.600              & 0.660                     & 0.575                    & 0.746                  & 0.542                      \\ \hline pbeans-2.0        & 0.738              & 0.745                 & 0.740                & 0.719                     & 0.730                  & 0.730              & 0.740                      & 0.736                      & 0.742                    & 0.734                  & 0.732                    & 0.617                     & 0.750              & 0.572                     & 0.802                    & 0.637                  & 0.738                      \\ \hline poi-1.5           & 0.756              & 0.773                 & 0.701                & 0.699                     & 0.705                  & 0.705              & 0.741                      & 0.644                      & 0.473                    & 0.709                  & 0.629                    & 0.764                     & 0.521              & 0.531                     & 0.595                    & 0.520                  & 0.603                      \\ \hline poi-2.0RC1        & 0.676              & 0.629                 & 0.595                & 0.674                     & 0.595                  & 0.595              & 0.620                      & 0.630                      & 0.678                    & 0.714                  & 0.581                    & 0.596                     & 0.663              & 0.566                     & 0.606                    & 0.641                  & 0.653                      \\ \hline poi-2.5.1         & 0.726              & 0.751                 & 0.777                & 0.725                     & 0.795                  & 0.795              & 0.695                      & 0.730                      & 0.793                    & 0.657                  & 0.432                    & 0.820                     & 0.525              & 0.623                     & 0.623                    & 0.516                  & 0.563                      \\ \hline poi-3.0           & 0.786              & 0.795                 & 0.858                & 0.646                     & 0.766                  & 0.766              & 0.734                      & 0.684                      & 0.840                    & 0.736                  & 0.761                    & 0.821                     & 0.578              & 0.593                     & 0.548                    & 0.554                  & 0.498                      \\ \hline synapse-1.0       & 0.779              & 0.731                 & 0.796                & 0.751                     & 0.710                  & 0.710              & 0.638                      & 0.615                      & 0.795                    & 0.741                  & 0.674                    & 0.763                     & 0.572              & 0.659                     & 0.410                    & 0.554                  & 0.667                      \\ \hline synapse-1.1       & 0.641              & 0.611                 & 0.607                & 0.671                     & 0.528                  & 0.528              & 0.658                      & 0.525                      & 0.668                    & 0.695                  & 0.543                    & 0.656                     & 0.589              & 0.614                     & 0.571                    & 0.493                  & 0.538                      \\ \hline synapse-1.2       & 0.709              & 0.682                 & 0.610                & 0.684                     & 0.667                  & 0.667              & 0.620                      & 0.677                      & 0.753                    & 0.625                  & 0.640                    & 0.711                     & 0.585              & 0.579                     & 0.612                    & 0.509                  & 0.523                      \\ \hline tomcat-6.0.389418 & 0.807              & 0.787                 & 0.771                & 0.753                     & 0.783                  & 0.783              & 0.772                      & 0.771                      & 0.601                    & 0.788                  & 0.649                    & 0.810                     & 0.476              & 0.377                     & 0.521                    & 0.646                  & 0.536                      \\ \hline velocity-1.4      & 0.525              & 0.603                 & 0.618                & 0.609                     & 0.584                  & 0.584              & 0.546                      & 0.522                      & 0.524                    & 0.504                  & 0.666                    & 0.450                     & 0.549              & 0.577                     & 0.484                    & 0.604                  & 0.478                      \\ \hline velocity-1.5      & 0.651              & 0.616                 & 0.642                & 0.612                     & 0.547                  & 0.547              & 0.622                      & 0.544                      & 0.391                    & 0.561                  & 0.629                    & 0.678                     & 0.577              & 0.600                     & 0.643                    & 0.496                  & 0.571                      \\ \hline velocity-1.6.1    & 0.706              & 0.672                 & 0.690                & 0.707                     & 0.589                  & 0.589              & 0.669                      & 0.544                      & 0.579                    & 0.657                  & 0.634                    & 0.611                     & 0.562              & 0.553                     & 0.560                    & 0.566                  & 0.545                      \\ \hline xalan-2.4.0       & 0.744              & 0.754                 & 0.792                & 0.735                     & 0.711                  & 0.711              & 0.756                      & 0.720                      & 0.755                    & 0.709                  & 0.672                    & 0.730                     & 0.628              & 0.626                     & 0.570                    & 0.724                  & 0.680                      \\ \hline xalan-2.5.0       & 0.637              & 0.674                 & 0.680                & 0.673                     & 0.629                  & 0.629              & 0.626                      & 0.631                      & 0.635                    & 0.644                  & 0.612                    & 0.688                     & 0.583              & 0.568                     & 0.590                    & 0.568                  & 0.545                      \\ \hline xalan-2.6.0       & 0.719              & 0.746                 & 0.728                & 0.701                     & 0.680                  & 0.680              & 0.717                      & 0.695                      & 0.708                    & 0.634                  & 0.666                    & 0.693                     & 0.631              & 0.553                     & 0.594                    & 0.627                  & 0.598                      \\ \hline xalan-2.7.0       & 0.844              & 0.847                 & 0.832                & 0.854                     & 0.644                  & 0.644              & 0.841                      & 0.640                      & 0.693                    & 0.787                  & 0.850                    & 0.817                     & 0.614              & 0.657                     & 0.634                    & 0.637                  & 0.619                      \\ \hline xerces-1.2.0      & 0.473              & 0.509                 & 0.490                & 0.579                     & 0.500                  & 0.500              & 0.475                      & 0.509                      & 0.458                    & 0.478                  & 0.583                    & 0.445                     & 0.573              & 0.562                     & 0.557                    & 0.522                  & 0.534                      \\ \hline xerces-1.3.0      & 0.651              & 0.663                 & 0.619                & 0.636                     & 0.664                  & 0.664              & 0.608                      & 0.609                      & 0.660                    & 0.614                  & 0.582                    & 0.656                     & 0.589              & 0.699                     & 0.603                    & 0.480                  & 0.466                      \\ \hline xerces-1.4.4      & 0.810              & 0.739                 & 0.715                & 0.683                     & 0.720                  & 0.720              & 0.760                      & 0.700                      & 0.795                    & 0.717                  & 0.741                    & 0.784                     & 0.558              & 0.535                     & 0.658                    & 0.489                  & 0.553                      \\ \hline xerces-init       & 0.603              & 0.607                 & 0.682                & 0.667                     & 0.579                  & 0.579              & 0.541                      & 0.593                      & 0.582                    & 0.473                  & 0.556                    & 0.525                     & 0.591              & 0.571                     & 0.591                    & 0.634                  & 0.632                      \\ \hline Mean              & 0.739              & 0.737                 & 0.734                & 0.729                     & 0.717                  & 0.717              & 0.714                      & 0.706                      & 0.702                    & 0.693                  & 0.688                    & 0.680                     & 0.639              & 0.636                     & 0.633                    & 0.624                  & 0.602                      \\ \hline SD                & 0.091              & 0.097                 & 0.095                & 0.084                     & 0.099                  & 0.099              & 0.095                      & 0.105                      & 0.129                    & 0.096                  & 0.098                    & 0.133                     & 0.089              & 0.097                     & 0.102                    & 0.095                  & 0.082                      \\ \hline Median            & 0.738              & 0.751                 & 0.742                & 0.719                     & 0.715                  & 0.715              & 0.724                      & 0.700                      & 0.701                    & 0.701                  & 0.669                    & 0.688                     & 0.613              & 0.629                     & 0.606                    & 0.630                  & 0.598                      \\ \hline \end{tabular} }}
\end{table*}

\end{document}